\documentclass{article}
\usepackage[utf8]{inputenc}
\usepackage{authblk}
\usepackage{setspace}
\usepackage[margin=1.25in]{geometry}
\usepackage{graphicx}
\graphicspath{ {./figures} }
\usepackage{subcaption}
\usepackage{amsmath}
\usepackage{listings}
\usepackage{threeparttable}
\usepackage{makecell} 
\usepackage{svg}

\usepackage[style=nejm, citestyle=numeric-comp,sorting=none]{biblatex}
\title{Deep Learning and Methods Based on Large Language Models Stellar Light Curve Classification}

\author[1,2,3]{Yu-Yang Li}
\author[1,2,3]{Yu Bai}
\author[1,2*]{Cunshi Wang}
\author[7,8]{Mengwei Qu}
\author[9]{Ziteng Lu}
\author[2,4,5]{Roberto Soria}
\author[1,2,3,6]{Jifeng Liu}

\affil[1]{Key Laboratory of Optical Astronomy, National Astronomical Observatories, Chinese Academy of Sciences,
20A Datun Road, Chaoyang District, Beijing 100101, People’s Republic of China}
\affil[2]{College of Astronomy and Space Sciences, University of Chinese Academy of Sciences, Beijing 100049, China}
\affil[3]{Institute for Frontiers in Astronomy and Astrophysics, Beijing Normal University, Beijing 102206, China}
\affil[4]{INAF—Osservatorio Astrofisico di Torino, Strada Osservatorio 20, I-10025 Pino Torinese, Italy}
\affil[5]{Sydney Institute for Astronomy, School of Physics A28, The University of Sydney, Sydney, NSW 2006, Australia}
\affil[6]{New Cornerstone Science Laboratory, National Astronomical Observatories, Chinese Academy of Sciences, Beijing 100101, People’s Republic of China}
\affil[7]{State Key Laboratory of Isotope Geochemistry, Guangzhou Institute of Geochemistry, Chinese Academy of Sciences, Guangzhou 510640,China}
\affil[8]{College of Earth and Planetary Sciences, University of Chinese Academy of Sciences, Beijing 100049, China}
\affil[9]{School of Foreign Studies, TonglingUniversity, Tongling, Anhui, 244061, People’s Republic of China}
\affil[*]{Address correspondence to: wangcunshi@nao.cas.cn}

\date{}

\begin{document}

\maketitle

\begin{abstract}

Light curves serve as a valuable source of information on stellar formation and evolution. With the rapid advancement of machine learning techniques, they can be effectively processed to extract astronomical patterns and information. In this study, we present a comprehensive evaluation of models based on deep learning and large language models (LLMs) for the automatic classification of variable star light curves, using large datasets from the Kepler and K2 missions. Special emphasis is placed on Cepheids, RR Lyrae, and eclipsing binaries, examining the influence of observational cadence and phase distribution on classification precision. Employing automated deep learning optimization, we achieve striking performance 2 architectures: one that combines 1-dimensional convolution (Conv1D) with bidirectional long short-term memory (BiLSTM) and another called the Swin Transformer. These achieved accuracies of 94\% and 99\% respectively, with the latter demonstrating a notable 83\% accuracy in discerning the elusive Type II Cepheids that comprise merely 0.02\% of the total dataset. We unveil StarWhisper LightCurve (LC), a series of 3 LLM models based on an LLM, a multimodal large language model (MLLM), and a large audio language model (LALM). Each model is fine-tuned with strategic prompt engineering and customized training methods to explore the emergent abilities of these models for astronomical data. Remarkably, StarWhisper LC series models exhibit high accuracies around 90\%, considerably reducing the need for explicit feature engineering, thereby paving the way for streamlined parallel data processing and the progression of multifaceted multimodal models in astronomical applications. The study furnishes 2 detailed catalogs illustrating the impacts of phase and sampling intervals on deep learning classification accuracy, showing that a substantial decrease of up to 14\% in observation duration and 21\% in sampling points can be realized without compromising accuracy by more than 10\%.  

\textbf{Keywords:} Computational astronomy, Variable stars, Large language model, Deep learning
\end{abstract}

\section*{Introduction} \label{sec:intro}
The phenomenon of light variation is a crucial aspect of astrophysics and has long been studied in the field of time-domain astronomy. Cepheid variables, a special kind of variable star, serve a critical role as a universal standard candle, enabling us to measure the distance to clusters and galaxies with their period-luminosity relation. Other types of variable stars, including RR Lyrae (RR), $\delta$ Sct (DSCT), and $\gamma$ Dor (GDOR) stars, are characterized by their unique absolute magnitudes, pulsation patterns, each exhibiting diverse period-luminosity relations. The classification of these pulsating stars has considerably enriched our understanding of their formation and evolution. Moreover, it has shed light on the structure and dynamics of binary, n-body systems, and our galaxy. \cite{prsaKEPLERECLIPSINGBINARY2011, slawsonKEPLERECLIPSINGBINARY2011, matijevicKEPLERECLIPSINGBINARY2012, conroyKEPLERECLIPSINGBINARY2014a, conroyKeplerEclipsingBinary2014b, lacourseKeplerEclipsingBinary2015, kirkKEPLERECLIPSINGBINARY2016, abdul-masihKEPLERECLIPSINGBINARY2016a}

Long periods of observations are crucial in order to understand the nature of variable stars. To carry out these observations, it is important to make wise use of telescope time, which is a valuable resource \cite{garcia-piquerEfficientSchedulingAstronomical2017}. One strategy for doing so involves the use of simulations to create plans that optimally allocate available telescope time, considering overheads and other factors. Another strategy employs autonomous agents to optimize the observation of time-varying phenomena \cite{saundersOptimalObservingAstronomical}. In this study, we aim to understand the importance tied to different phases of variables and the classification accuracy provided by varying cadences. This can help us to establish robust guidelines for future star gazing observations \cite{wangTransferLearningApplied2023}.

A substantial amount of labeled data is required, in order to understand the significance of phase and cadence for different variable stars. In recent years, astronomy has entered the era of big data. For example, the Zwicky Transient Facility (ZTF) \cite{bellmZwickyTransientFacility2014} at Palomar Observatory scans the sky every 2 days with a 1.2m telescope , generating approximately 1 TB of raw data each night \cite{mahabalMachineLearningZwicky2019a}. The upcoming SiTian survey  \cite{liuSiTianProject2021}, aims to monitor over 1,000 $deg^2$ of sky every 30 minutes using fifty 1m class Schmidt telescopes, and is expected to produce around 140TB of processed data each night.

As the volume of data grows, there is an increasing requirement for efficient and automated interpretation and analysis methods. Deep learning, a component of machine learning, has become a powerful tool for image and signal processing \cite{lecunDeepLearning2015}. It can learn and extract features from data \cite{krizhevskyImageNetClassificationDeep2012a,lecunDeepLearning2015}, making it an ideal solution for managing large, complex datasets. Specifically, recurrent neural networks (RNNs) have shown great effectiveness in processing time series data \cite{liptonCriticalReviewRecurrent2015}, while convolutional neural networks (CNNs) have demonstrated their superiority in image processing tasks \cite{krizhevskyImageNetClassificationDeep2012a,heDeepResidualLearning2015}. Moreover, the transformer architecture, with its attention mechanism \cite{vaswaniAttentionAllYou2017}, has shown remarkable performance across various applications \cite{devlinBERTPretrainingDeep2019a,radfordImprovingLanguageUnderstanding,radfordLanguageModelsAre, brownLanguageModelsAre2020}. RNNs originally developed by \cite{elmanFindingStructureTime1990}, led to Simple RNNs, which were later enhanced by long short-term memory (LSTM,  \cite{hochreiterLongShortTermMemory1997}) to address long-term dependencies. These were further simplified by gated recurrent units (GRUs) \cite{choPropertiesNeuralMachine2014} for computational efficiency. CNNs initially introduced for hand written digit recognition
 \cite{lecunBackpropagationAppliedHandwritten1989}, and have evolved with architectures such as AlexNet, VGGNet, and ResNet. Recent advancements include EfficientNet  \cite{tanEfficientNetRethinkingModel2020, tanEfficientNetV2SmallerModels2021}, which optimizes depth, width, and resolution for model accuracy and efficiency.

In recent years, transformer architecture has shown remarkable performance across a diverse range of applications.Initially introduced to address sequential data processing tasks \cite{vaswaniAttentionAllYou2017}. Transformers use self-attention mechanisms to model relationships with sequences, enabling them to capture long-range dependencies and relationships between sequence elements. The success of the transformer in natural language processing has sparked interest in their potential application in other domains, such as computer vision  and speech recognition. An example of a vision transformer is the Swin Transformer \cite{liuSwinTransformerHierarchical2021,liuSwinTransformerV22022}, which has optimized the computation of the attention mechanism and shown promising results in various benchmark datasets.

Beyond the traditional reliance on extensive data for transfer learning, it is imperative to investigate the potential of large language models (LLMs) that incorporate both substantial datasets and large numbers of parameter. StarWhisper (available at https://github.com/Yu-Yang-Li/StarWhisper) is an LLM for astronomy that has strong astronomical ability and instruction-following ability and can complete a series of functions such as knowledge question answering, calling multimodal tools, and docking telescope control systems. The StarWhisper LC Series is initiated with the purpose of leveraging the experience gained from the prior training of the StarWhisper language model. It seeks to explore and discuss the potential harnessed from vast data to engender emergent properties in the analysis of light curve data. LLMs such as the Gemini 7B model have shown promise in adapting to new data types through fine-tuning with specific prompt templates \cite{zhou2023fits13}. Multimodal large language models (MLLMs) such as the deepseek-vl-7b-chat, are adept at handling tasks involving image classification due to their extensive training on datasets containing chart data \cite{tsai2019multimodal_6}. Large audio language models (LALMs), such as Qwen-audio,trained on audio datasets, exhibit exceptional performance in audio classification \cite{yang2022voice2series_4}.

In this study, we perform a comprehensive evaluation of deep-learning and LLM-based models for the classification of variable star light curves, using the data from Kepler and K2 observations. The types of variable stars and the data pre-processing are presented in the section on data. In the section on models, we utilize various classification models, including LSTM, GRU, transformer, LightGBM, EfficientNet, the Swin Transformer and the StarWhisper LC series. Then we introduce training methods. We also present model performance and catalogs of the phase importance and sampling intervals in the section on results. A discussion is provided in the discussion section.

\section*{Materials and Methods} \label{sec:sec2}
\subsection*{Data} \label{sec:data}

\subsubsection*\paragraph{Kepler and K2}

The Kepler spacecraft was launched in 2009 aiming to discover Earth like planets \cite{doi:10.1126/science.1185402}. It was equipped with an optical telescope with a 95 cm aperture and a $115.6^\circ$ field of view. The advanced technology allowed Kepler to precisely track light curves from 200,000 different targets. This high precision resulted in the discovery of over 2,000 planets. For stars with V band magnitude between 13 mag and 14 mag, the precision was 100 parts per million (ppm), while for stars with V band magnitude between 9 mag to 10 mag, the precision was 10 ppm. Unfortunately, after 4 years, half of Kepler's 4 reaction wheels failed, leading to the end of its primary mission. Yet, this marked the beginning of the K2 mission. The K2 mission used the transit method to detect changes in light along the ecliptic plane and created catalogs with photometric precision closely matching that of the original Kepler mission. The observations of K2 were controlled using the remaining reaction wheels and thrusters, with each campaign limited to 80 days.

There are 2 types of light curves available: presearch data conditioning (PDC) and simple aperture photometry (SAP) light curves \cite{Cleve2016KeplerIH}. SAP light curves preserve long-term trends, whereas PDC light curves, produced by the Kepler Operations Center, are corrected for systematic errors. Consequently, we opt for PDC light curves in our analysis. Prior to testing, we evaluated 2 data types with distinct temporal resolutions: long-cadence data, sampled every 30 minutes, and short-cadence data, sampled every minute. Due to the limited availability of short-cadence data, we decided not to use it and focused exclusively on long-cadence light curves. For each light curve, we eliminate quarter-to-quarter variations and normalize the flux to relative flux, following the methodologies outlined in \cite{yangFlareCatalogFlare2019, hanStellarActivityCycles2021}.

\subsubsection*{Variable stars} \label{subsec:variable}

Our training samples are similar to \cite{wangTransferLearningApplied2023}, including eclipsing binaries (EBs), RRs, DSCT, GDORs, and DSCT / GDOR hybrids (HYBs). The type II Cepheids (T2CEP) are included, in order to make a more universal sample. Table \ref{tab:sample} lists our training samples and corresponding references. The training samples are seriously biased among different variable types. This bias often leads to decreased performance, especially in few-shot situations and small sample scenarios \cite{kokolMachineLearningSmall2022,clemenconStatisticalLearningBiased2019}. However, \cite{taniguchiMachineLearningModel2018} suggest that certain models may help overcome these challenges, although such cases seem to be the exception rather than the rule. These biased training samples offer a unique opportunity to study the algorithm's dependence on such biases and to understand how this dependence affects the overall performance and accuracy when applied to astronomical data. 

\subsubsection*{Pre-processing} \label{subsec:pre}

A pre-processing method, as described by \cite{wangTransferLearningApplied2023}, was adopted to manage and clean the light curves, with the aim of enhancing their features and expanding the training data. The lightcurves were segmented into 10-day intervals. Any segments with gaps exceeding 1 day were removed, while those with gaps less than 1 day were interpolated with a time sequence of 30-minute intervals.

\begin{table*}[!t]
    \centering
    \caption{Training Samples}
    \label{tab:sample}
    \begin{threeparttable}
        \begin{tabular}{lccc}
        \hline
        \hline
            Label        & Input Catalog        &  Final Samples        &  References\\ \hline
            DSCT & 1,389          & 111,528          & \cite{kholopovCombinedGeneralCatalogue1998}, \cite{durlevichListErrorsGCVS1994}, \cite{artyukhinaVizieROnlineData1996}, \cite{bradleyRESULTSSEARCHDOR2015}, \cite{balonaGaiaLuminositiesPulsating2018}       \\
            EB           & 2,908          & 226,937          & \cite{prsaKEPLERECLIPSINGBINARY2011}, \cite{slawsonKEPLERECLIPSINGBINARY2011}, \cite{matijevicKEPLERECLIPSINGBINARY2012}, \cite{conroyKEPLERECLIPSINGBINARY2014a} , \cite{conroyKeplerEclipsingBinary2014b}, \cite{lacourseKeplerEclipsingBinary2015}, \cite{kirkKEPLERECLIPSINGBINARY2016}, \cite{abdul-masihKEPLERECLIPSINGBINARY2016}       \\
            GDOR & 941           & 65,786           & \cite{bradleyRESULTSSEARCHDOR2015}, \cite{balonaGaiaLuminositiesPulsating2018}        \\
            HYB          & 1,552          & 33,751           & \cite{kholopovCombinedGeneralCatalogue1998}, \cite{durlevichListErrorsGCVS1994}, \cite{artyukhinaVizieROnlineData1996}, \cite{samus84thNameListVariable2021}, \cite{bradleyRESULTSSEARCHDOR2015}, \cite{balonaGaiaLuminositiesPulsating2018}        \\ 
            RR           & 482           & 9,306            & \cite{molnarGaiaDataRelease2018}        \\
            T2CEP        & 3             & 94              & \cite{kholopovCombinedGeneralCatalogue1998}, \cite{durlevichListErrorsGCVS1994}, \cite{artyukhinaVizieROnlineData1996}, \cite{samus84thNameListVariable2021}, \cite{molnarGaiaDataRelease2018}        \\\hline
            Total        & 7,275          & 447,402          &         \\\hline
        \end{tabular}
        \begin{tablenotes}
            \small
            \item[\textnormal{Note}] The input catalog column indicates the sources of data that we obtained from several catalogs. The final sample column represents the final sample size after applying all the preprocessing procedures.  
        \end{tablenotes}
    \end{threeparttable}
\end{table*}

\begin{table*}[!t]
    \centering
    \caption{Period and Observation Time Saving}
    \label{tab:Save}
    \begin{threeparttable}
        \begin{tabular}{lcccccc}
        \hline
        \hline
            \textbf{Star} & \textbf{Period (d)} & \textbf{$\Delta_{phase}$ (\%)} & \textbf{$t_{phase}$ (\%)} & \textbf{$\Delta_{sampling}$ (\%)} & \textbf{$t_{sampling}$ (\%)} \\
        \hline
            RR\_251457011 & 0.509 & 1.06 & 41 & 8.63 & 50 \\
            RR\_251457012 & 0.529 & 3.48 & 71 & 0 & 0 \\
            RR\_251457013 & 0.542 & 2.47 & 31 & 1.83 & 50 \\
            RR\_251457014 & 0.593 & 9.99 & 29 & 9.32 & 50 \\
            RR\_251457015 & 0.574 & 0.53 & 55 & 7.82 & 50 \\
            RR\_251457016 & 0.510 & 1.30 & 83 & 1.38 & 50 \\
            RR\_251457020 & 0.478 & 0 & 0 & 0.76 & 50 \\
            RR\_251457021 & 0.546 & 0.70 & 74 & 4.48 & 80 \\
            RR\_251457022 & 0.508 & 1.56 & 49 & 0 & 0 \\
            RR\_251457024 & 0.553 & 0 & 0 & 0 & 0 \\
        \hline
        \end{tabular}
        \begin{tablenotes}
            \small
            \item[\textnormal{Note}] The table shows the period of the stars (in days) along with the variation of accuracy and time saved (percentages), for both phase importance and sampling research.
        \end{tablenotes}
    \end{threeparttable}
\end{table*}

An alternative approach to time-series data classification involves the categorizing of images that are created from graphically represented light curves. This is achieved using transfer learning techniques. As highlighted in their result, the continuous wavelet transform (CWT) method has demonstrated superior results in imaging light curves.

\begin{figure}[ht!]
    \centering
    \includegraphics[width=0.9\textwidth]{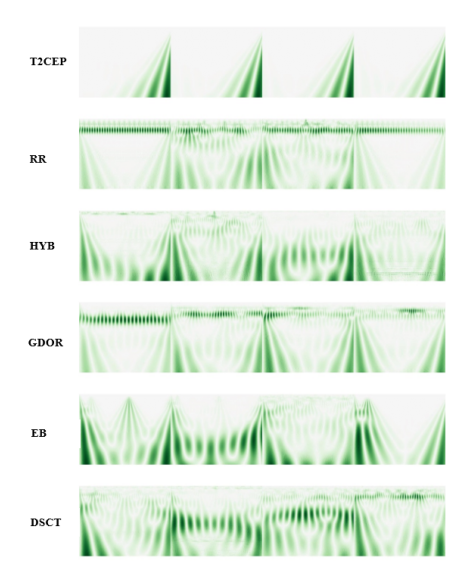}
    \caption{CWT images of different objects. Each row contains 4 objects, each labeled differently, showcasing their distinct features through CWT images.}
    \label{fig:cwt}
\end{figure}

The Morlet wavelet was chosen as the core function for the CWT, owing to its proven effectiveness in analyzing signals that display shifts in amplitude. This analysis resulted in a collection of images that provide valuable insights into the time-frequency characteristics of the signals. Importantly, these images facilitate the identification of patterns, trends, and anomalies, thus offering a comprehensive understanding of the signals. 

Fig. \ref{fig:cwt} shows some variable stars from the training sample. The CWT image displays a time-frequency diagram, with the highest frequencies at the top of the image. T2CEP stars have the longest periods, so the top and middle parts of the CWT image are empty \cite{percy2011Understanding}. RR stars pulsate with a very stable period of about 0.5 days and consistent amplitude, resulting in clear features at the top of the CWT images. HYB stars exhibit multiple types of variability, leading to features throughout the CWT images. DSCTs have multiple pulsation modes with periods ranging from 0.03 to 0.3 days \cite{dup05DSCT}, which are shorter than the pulsation periods of the GDORs, which range from 0.3 to 3 days. The CWT images of DSCTs show higher-frequency features compared to those of GDOR stars, and also shows different pulsation modes. For EBs, the light variations are caused by the eclipses of the 2 stars, resulting in a single clear feature in the CWT image. The periods of EBs vary, and some may have longer eclipse durations, represented by deeper straight bars.

\subsubsection*{Lomb-Scargle periodogram}

\begin{figure*}[ht!]
\centering
\includegraphics[width=0.9\textwidth]{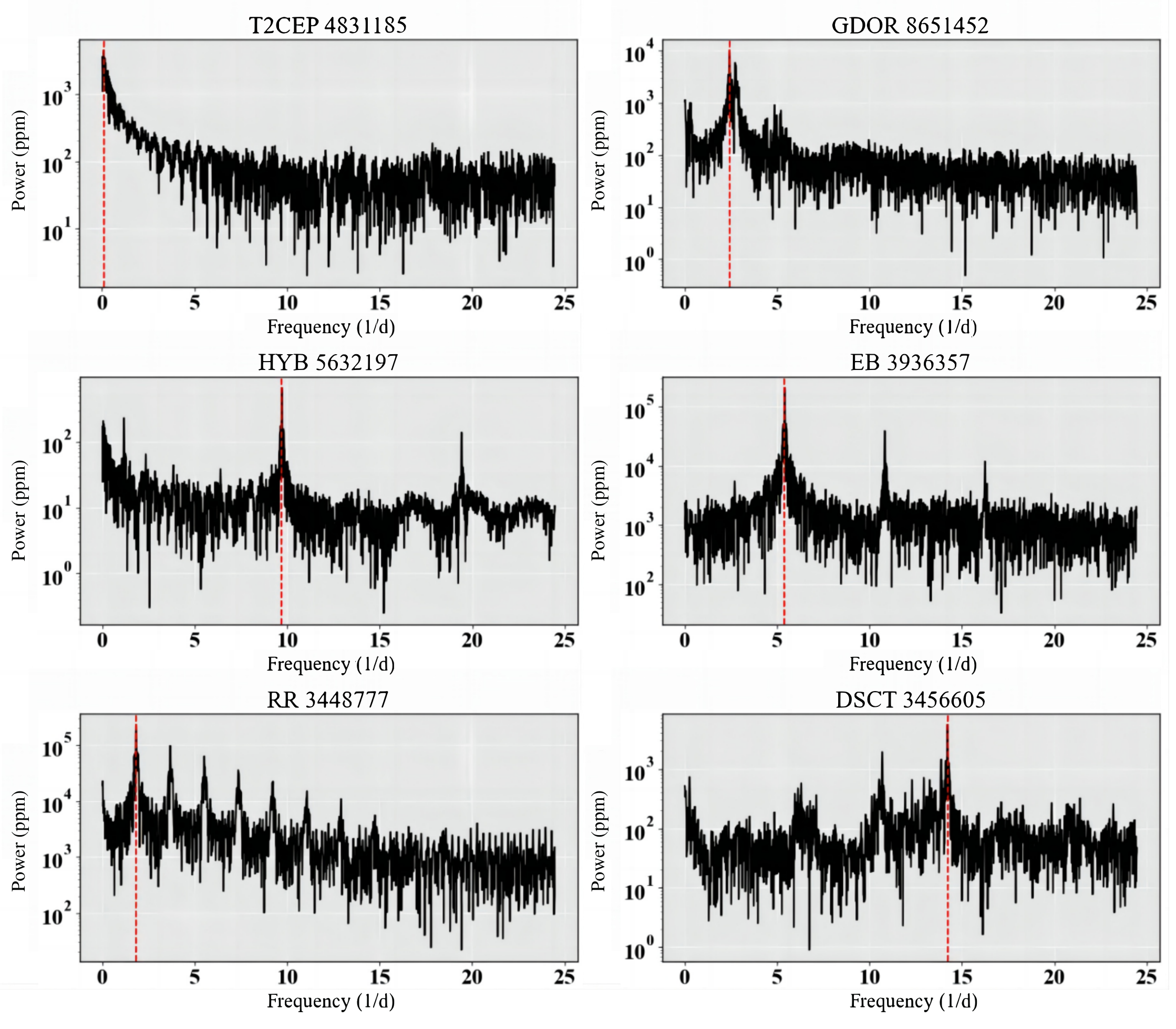}
\caption{Images of different variable types processed using the Lomb-Scargle method, with a selected frequency range
from 0 to Nyquist frequency. The final frequency is also indicated in each figure.
\label{fig:LS}}
\end{figure*}

We employ the Lomb-Scargle algorithm to extract the most significant periods complete light curves, which are used to study the phase importance of periodic variables. The Lomb-Scargle algorithm, a variant of the discrete fourier transform (DFT) developed by \cite{1976Ap&SS..39..447L} and \cite{scargleStudiesAstronomicalTime1982}, has been specifically designed for unevenly sampled time-series data. It transforms a time series into a linear combination of sinusoidal waveforms, simplifying the conversion from the time domain to the frequency domain. We calculate the period by applying the Lightkurve Collaboration's method \cite{vanderplasUnderstandingLombScarglePeriodogram2018}. Some examples are shown in Figure \ref{fig:LS}.

More specifically, we use a uniform sampling strategy in our set frequency domain to select an array of frequency points for periodicity analysis. The periodogram calculation is carried out by assessing the power spectral density (PSD) at each frequency. This periodogram reveals the intensity of periodic signals across the spectrum of frequencies. Significant peaks in the periodogram indicate strong periodic signals. We determine the exact periods through the reciprocals of these frequencies. Our implementation also includes considerations for normalization methods and computational strategies. We use a specific approach to frequency sampling to ensure that the periodogram computation is both precise and efficient. The results are shown in Table \ref{tab:Save}.

To assess the statistical significance of detected periods, we calculate the False Alarm Probability (FAP) for each period. The FAP is computed using the method provided by the LombScargle object in the Lightkurve library \cite{2018ascl.soft12013L}. This method evaluates the probability of observing a signal of equal or greater power in random noise data. We adopt a threshold of FAP \textless 0.001 to identify statistically significant periods. This criterion indicates that there is less than a 0.1\% chance that a detected period could arise from random fluctuations in the data, thus providing a rigorous basis for distinguishing genuine periodic signals from noise.

\subsection*{Model construction} \label{sec:model}

\begin{figure*}[ht!]
\centering
\includegraphics[width=0.9\textwidth]{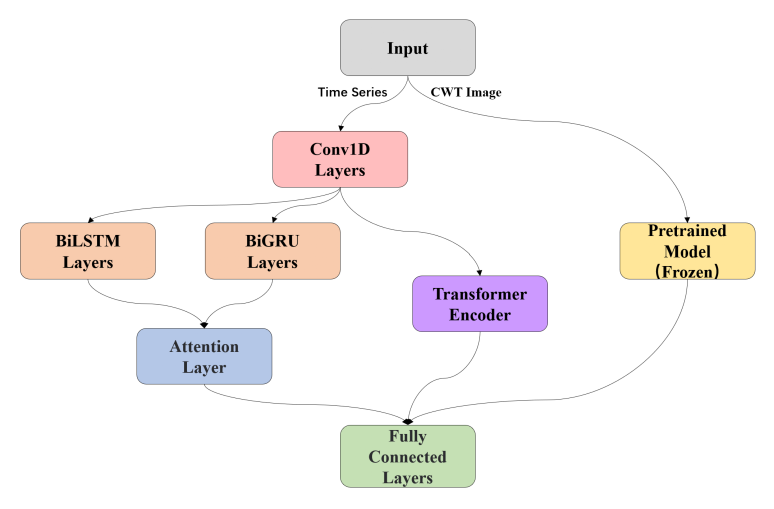}
\caption{Deep learning model structures for different input formats. Not
all branch combination models are adopted due to varying degrees of effectiveness. LLM models are not shown.
\label{fig:model}}
\end{figure*}

\begin{figure*}[ht!]
\centering
\includegraphics[width=0.9\textwidth]{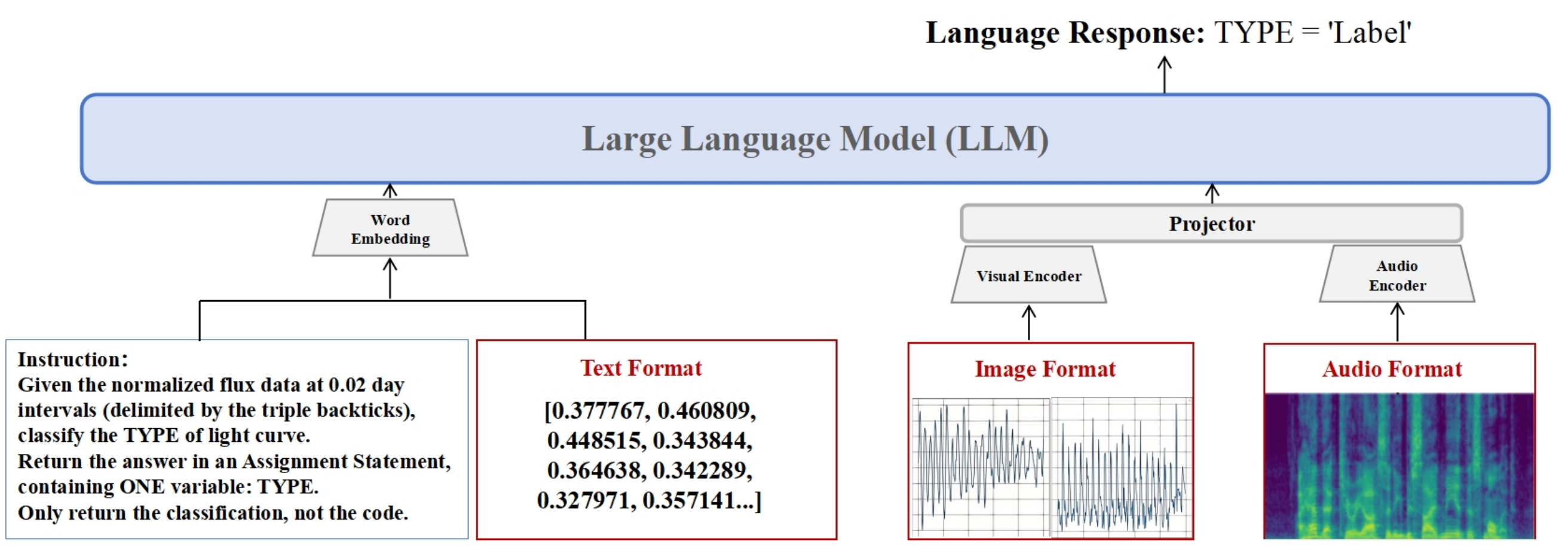}
\caption{Illustration of the LLM-based architecture of the StarWhisper LC series.
\label{fig:LLM-based}}
\end{figure*}

In order to classify variable stars from their light curves, we have explored various deep learning models. The basic architectures of these models are shown in Figure \ref{fig:model} and \ref{fig:LLM-based}. Our model selection was guided by the specific characteristics of variable star light curves and the unique advantages offered by each architecture.

CNN and RNN models, like 1-dimensional convolutional layers (Conv1D) combined with bidirectional long short-term memory (BiLSTM), were selected due to their proficiency in capturing temporal dependencies and local characteristics in time series data \cite{schusterBidirectionalRecurrentNeural1997,kimConvolutionalNeuralNetworks2014}. The bidirectional aspect of BiLSTM aids in identifying long-term dependencies, whereas Conv1D layers excel at extracting local features, thus enhancing the overall context. Transformer models, characterized by their self-attention mechanism \cite{vaswaniAttentionAllYou2017}, stand out in managing long sequences and capturing global dependencies, making them ideally suited for analyzing variable star data with intricate periodic patterns. In our image-based methods employing CWT, we utilized the Swin Transformer \cite{liuSwinTransformerHierarchical2021,liuSwinTransformerV22022} and EfficientNet \cite{tanEfficientNetRethinkingModel2020,tanEfficientNetV2SmallerModels2021}. The hierarchical windowed attention mechanism of the Swin Transformer is adept at capturing multi-scale features, which is particularly beneficial in few-shot learning contexts. Meanwhile, EfficientNet achieves an optimal balance between network depth, width, and resolution to ensure efficient feature extraction. Furthermore, we investigated the application of large language models via the StarWhisper LC series (LLM, MLLM, LALM), harnessing their robust generalization abilities and adaptability to novel tasks in time series data classification.

\subsubsection*{CNNs and RNNs} \label{subsec:lstm}

We explore several advanced LSTM architectures. Conv1D + BiLSTM combines 1-dimensional convolutional layers with bidirectional LSTM layers. The convolutional layer is used to extract features from the data \cite{kimConvolutionalNeuralNetworks2014}, while the bidirectional LSTM layer captures contextual information from both past and future time steps \cite{schusterBidirectionalRecurrentNeural1997}. The attention mechanism of the Conv1D + BiLSTM + Attention model enables it to focus on specific parts of the input data, improving performance by giving higher weights to more relevant parts \cite{vaswaniAttentionAllYou2017}. This can improve the accuracy of classification and facilitate further research on the importance of different phases for classification \cite{salinasDistinguishingPlanetaryTransit2023}.

GRU models are a variation of LSTM that are computationally more efficient as they have fewer parameters, making them easier to train and less prone to overfitting \cite{choPropertiesNeuralMachine2014,chungEmpiricalEvaluationGated2014}. Specifically, we use the Conv1D + BiGRU architecture.

In addition to deep learning approaches, it is also important to consider the performance of classic machine learning methods for comparison. One such method is LightGBM, a gradient-boosting decision tree (GBDT) framework that uses tree-based learning algorithms \cite{keLightGBMHighlyEfficienta}.

\subsubsection*{Transformers} \label{subsec:transformer}

Transformers are highly effective in managing data dependencies, a critical aspect of time series analysis. This results in favorable outcomes \cite{kitaevReformerEfficientTransformer2019,liuPYRAFORMERLOWCOMPLEXITYPYRAMIDAL2022}. Their multiple attention mechanisms are skilled in identifying the most relevant parts of input data \cite{vaswaniAttentionAllYou2017}. By combining a 1-dimensional convolutional layer with a transformer encoder layer, we can effectively capture both global dependencies and local interactions in the data. This approach addresses the transformer architecture's limitation in detecting local nuances.

We have adopted the Swin Transformer for few-shot classification tasks (i.e. T2CEP). Transformers excel in modeling long-range dependencies in computer vision tasks \cite{dosovitskiyImageWorth16x162021a, wangPyramidVisionTransformer2021} by focusing directly on an image's essential parts \cite{vaswaniAttentionAllYou2017}. The Swin Transformer takes this a step further; it streamlines computation by limiting attention within small windows while maintaining effectiveness \cite{liuSwinTransformerHierarchical2021,liuSwinTransformerV22022}.

\begin{table*}[!t]
    \centering
    \caption{Accuracy, Precision, Recall, and Macro F1 Score}
    \label{tab:accuracy}
    \begin{threeparttable}
        \begin{tabular}{|l|l|l|l|l|l|}
        \hline
            Input Format & Model & \makecell{Accuracy\\ (All/T2CEP)} & Precision & Recall & \makecell{Macro\\ F1- Score} \\ \hline
            Time-Series & Conv1D + Transformer & 85\%/17\% & 0.69 & 0.76 & 0.70 \\ \cline{2-6}
            & LightGBM & 87\%/25\% & 0.70 & 0.81 & 0.71 \\ \cline{2-6}
            & BiLSTM + Attention & 93\% & 0.91 & 0.92 & 0.76 \\ \cline{2-6}
            & Conv1D + BiGRU & 93\% & 0.91 & 0.91 & 0.76 \\ \cline{2-6}
            & Conv1D + BiLSTM & 94\% & 0.92 & 0.94 & 0.77 \\ \hline
            CWT Image & EfficientNet & 99\% & 0.98 & 0.99 & 0.82 \\ \cline{2-6}
            & Swin Transformer & 99\%/83\% & 0.96 & 0.95 & 0.96 \\ \hline
            Textual Time-Series & LLM-based model & 89\% & 0.77 & 0.87 & 0.80 \\ \hline
            Light Curve Image  & MLLM-based model & 95\% & 0.93 & 0.95 & 0.94 \\ \hline
            Transformed Audio & LALM-based model & 93\% & 0.88 & 0.90 & 0.71 \\ \hline
        \end{tabular}
        \begin{tablenotes}
            \small
            \item[\textnormal{Note}] The table summarizes various performance metrics: accuracy represents the proportion of true results; precision indicates the relevancy of results; recall shows the model's ability to retrieve all relevant instances; and macro F1 score balances precision and recall across classes.
        \end{tablenotes}
    \end{threeparttable}
\end{table*}

\subsubsection*{EfficientNet}\label{subsec:efficientnet}

For a more thorough comparative analysis, we have adopted an advanced pretrained CNN known as EfficientNet. This network exceeds its predecessors by integrating optimization elements for depth, width, and resolution within its framework. In comparison, previous networks only optimized 1 or 2 of these elements \cite{tanEfficientNetRethinkingModel2020, tanEfficientNetV2SmallerModels2021}. EfficientNet uses a diverse range of architectural techniques including depth-wise separable convolutions, squeeze-and-excitation blocks, and dynamic image scaling. These techniques are validated to boost performance while reducing computational resource requirements.

\subsubsection*{StarWhisper LC series}

\emph{LLM}

LLMs, owing to their emergent capabilities, can learn new languages (broadly defined) through few-shot learning or fine-tuning processes, such as lean3 \cite{ying2024internlmmath}. The Gemini 7B model (which belongs to Google’s Gemini family, reported at:  \url{storage.googleapis.com/deepmind-media/gemini/gemini_1_report.pdf}), was pretrained on a vast corpus of text, and exhibits emergent behavior, enabling it to perform tasks not explicitly covered during its initial training. Therefore, we are contemplating the integration of specific prompt template as shown in \ref{tab:prompt-output}, to transform time series data into a form of language in a broader sense, in order to leverage these advanced models more effectively.

\emph{MLLM}

\begin{figure*}[ht!]
\centering
\includegraphics[width=\textwidth]{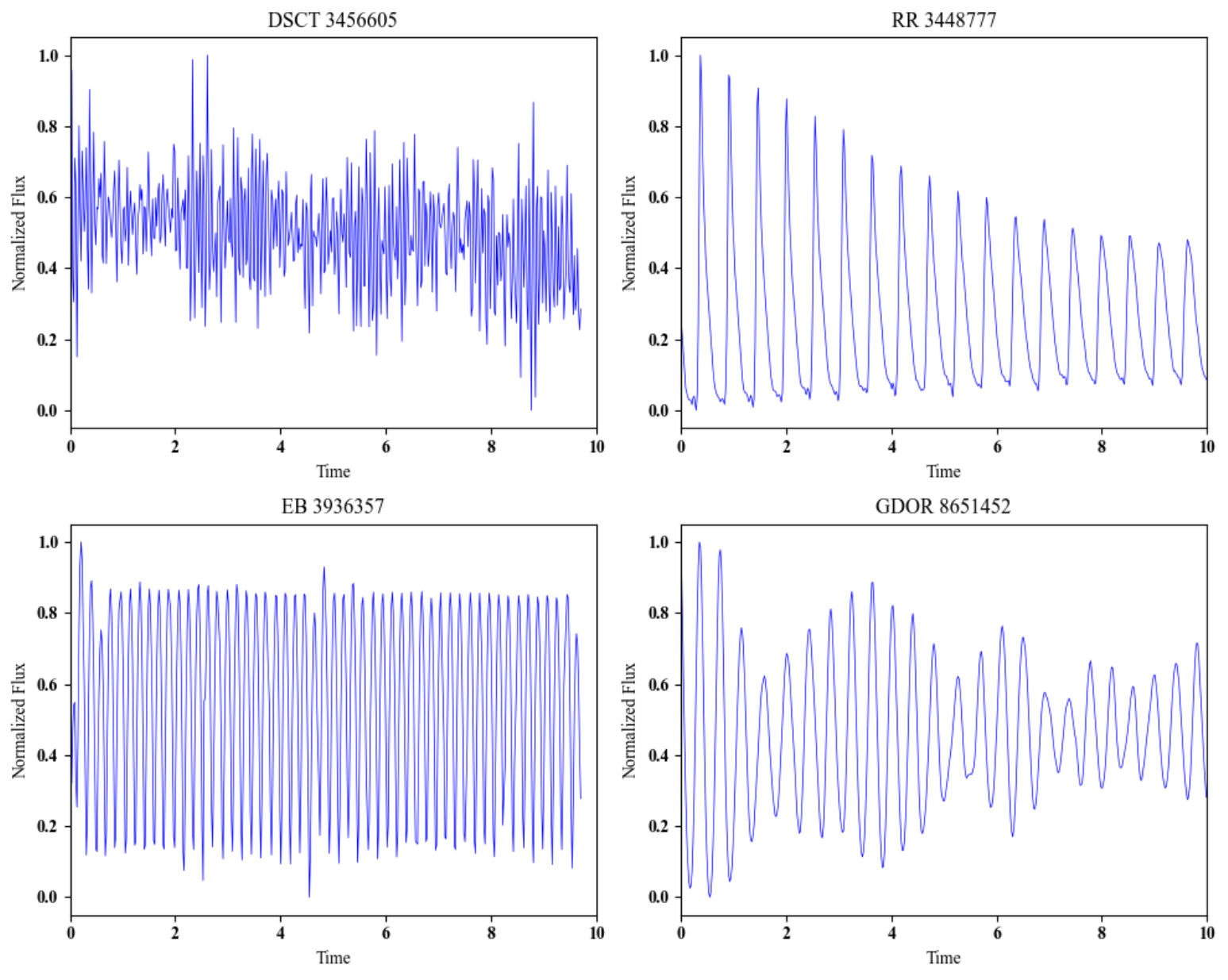}
\caption{Time series images of different objects.
\label{fig:image}}
\end{figure*}

Next, we introduce an MLLM trained on deepseek-vl-7b-chat \cite{lu2024deepseekvl}. MLLMs expand upon the capabilities of LLMs by incorporating the ability to process and interpret visual information. The deepseek-vl-7b-chat model, with its extensive training on chart data, is particularly well suited for the time series images such as those shown in Fig. \ref{fig:image}, making it a valuable addition to the StarWhisper LC series.

\emph{LALM}

The final model in the series is based on audio processing, utilizing Qwen-Audio \cite{chu2023qwenaudio}. This model has been specifically enhanced for audio classification tasks, offering a novel approach to time series analysis by converting time series data into audio signals. A sampling frequency of 500Hz is used to transform the normalized time-series data into audio signals, which possess distinct characteristics that aid in classification. Although higher frequencies are commonly employed for audio transfer (\url{https://www.mathworks.com/help/matlab/ref/audiowrite_zh_CN.html}), the resulting audio files can be too short for training the LALM. This approach may introduce distortion issues; however, it achieves higher accuracy and $\rm F_1$ scores compared to the Conv1D + Transformer method and the LightGBM method, demonstrating considerable promise.

\begin{table}[h]
\centering
\caption{Prompt and Output Template}
\label{tab:prompt-output}
\begin{threeparttable}
\begin{tabular}{p{\textwidth}}
\hline
\textbf{Instruction} \\
Given the normalized flux data at 0.02 day intervals (delimited by the triple backticks), classify the TYPE of light curve. 

Return the answer in an Assignment Statement, containing ONE variable: TYPE. 

Only return the classification, not the Python code.
\begin{verbatim}
```{flux}```
\end{verbatim} \\
\hline
\textbf{Output} \\
TYPE = `Label` \\
\hline
\end{tabular}
\begin{tablenotes}
\small
\item For other LLM-based models, replace “Python code” with “description,” and “flux” with “image path or audio path.”
\end{tablenotes}
\end{threeparttable}
\end{table}

\subsection*{Training} \label{sec:auto}

\subsubsection*{Automated deep learning methods}

Automated deep learning methods are techniques that aim to automate the process of designing, training, and evaluating neural network models. These methods typically use techniques such as Bayesian optimization \cite{snoekPracticalBayesianOptimization2012}, reinforcement learning \cite{mnihHumanlevelControlDeep2015}, or evolutionary algorithms \cite{realLargeScaleEvolutionImage2017} to optimize the model architecture and hyperparameters for the best possible performance.

Bayesian optimization is a powerful approach for hyperparameter tuning and exploration\cite{bergstraAlgorithmsHyperParameterOptimization2011}. It utilizes Gaussian processes regression (GPR) or tree-structured Parzen estimators (TPEs) to model the probability distribution of historical data and employs the sequential model-based optimization (SMBO) framework for iterative hyperparameter selection. 

The SMBO process is an iterative method for hyperparameter optimization that consists of 4 main steps \cite{10.1007/978-3-642-25566-3_40}. First, a probability distribution model is built based on the existing tuning history. Next, an acquisition function, such as expected improvement (EI), is employed to select the next hyperparameter. Following this, the new observation is incorporated into the existing tuning history. The process is then repeated until the predefined maximum number of iterations is reached.

In this study, we used automated deep learning to automate the process of training and evaluating deep learning models. Bayesian optimization allowed us to efficiently find the best set of hyperparameters, such as learning rate, batch size, number of network layers, and dropout rate. The implementation of a pruning mechanism enabled the premature termination of suboptimal trials, considerably enhancing the overall efficiency of the optimization process \cite{akibaOptunaNextgenerationHyperparameter2019}. We allocated 80$\%$ of the data for training and the remaining 20$\%$ for validation. All experiments were conducted on a computing system equipped with 2 NVIDIA A6000 GPUs.

\subsubsection*{LLM-based model training}

We faced considerable memory constraints that posed a substantial threat to our ability to effectively optimize our model. To mitigate these challenges, we first implemented the low-rank adaptation (LoRA) method, as outlined by \cite{hu2021LoRAlowrankadaptationlarge}. LoRA provides a memory-efficient solution for adapting pre-trained models to specific tasks through the use of low-rank matrix factorization. This technique minimizes the number of parameters that must be stored and updated during training, thus it enabled us to fine-tune our model under stringent memory conditions without compromising on performance, while also achieving a notable reduction in memory usage.

Despite the advantages of LoRA, there were instances where its optimized memory footprint still fell short of our needs. In response, we adopted quantized low-rank adaptation (QLoRA), an advanced variant of LoRA introduced by \cite{dettmers2023qLoRA}. QLoRA integrates quantization techniques to further reduce the precision of the model parameters. This additional step led to a more pronounced decrease in both memory requirements and computational overhead, allowing us to accommodate larger models within the confines of our available memory resources and maintain comparable performance levels.

The light curve data was prepared to fit within our LLM’s context length limitations and underwent a normalization process. The fine-tuning process involved all layers of the model using the QLoRA technique. The MLLM's training involved a comprehensive fine-tuning process that adjusted both the textual and visual processing components of the model. This approach was necessary to effectively handle the multimodal nature of the task, leveraging the model's pretrained knowledge of visual patterns for time series image classification. For the LALM, the success of the training phase hinged on the conversion of time series data into audio signals. This approach allowed the model to classify light curves transformed into audio signals, demonstrating the potential of audio-based analysis in scientific research.

\section*{Results and Discussion} \label{sec:sec3}
\subsection*{Results} \label{sec:result}

\subsubsection*{Performance}
 \label{subsec:confusion}

The confusion matrices (included in the Supplementary Materials) serve as the basis for computing numerous performance metrics. $\rm F_1$ score, a balance indicator, is calculated as the harmonic mean of precision and recall. We here calculate the Macro$-\rm F_1$ score \cite{DBLP:journals/corr/abs-1911-03347} to evaluate the model's overall performance, which averages the F1 scores across all classes, making it a vital measure when working with imbalanced data.

For transfer learning, both models achieved an accuracy of 99$\%$. The Swin Transformer achieved the highest Macro$-\rm F_1-$score of 0.96, demonstrating high effectiveness even with a small sample size. It achieved an 83$\%$ accuracy rate on T2CEP, a class comprising only 94 out of 447,402 samples. For the non-pretrained models, RNN-based models showed efficiency while maintaining high accuracy levels. The Conv1D+BiLSTM model showed the best performance with an accuracy of 94$\%$. It achieved a Macro-F1 score of 0.77, comparable to that of the pretrained EfficientNet. This proves its ability to manage imbalanced data.

Transfer learning's good performance is due to preprocessing via the CWT, a technique that simplifies the process of initial feature extraction. Its efficiency is further supported by its prior training on numerous images, enhancing its feature extraction capability. However, the HYB category performs short relative to others, when evaluating classification accuracy for variable stars. This underperformance could be due to the intrinsic complexity of the HYB category, resulting in an overlap of features and subsequently complicating the classification process \cite{10.1093/mnras/sty1511,wangTransferLearningApplied2023}.

The StarWhisper LC series also yielded impressive results in the classification of light curves, showcasing the robust capabilities of LLM-based models in scientific data analysis. The LLM, despite the considerable data reduction achieved by trimming the time-series data to input samples of 0.2d and normalizing the data's precision to one part in a hundred thousand, attained an accuracy rate of approximately 89\%. This demonstrates the model's ability to efficiently process and analyze condensed time-series data without substantial loss of information. The MLLM, which did not utilize CWT remarkably achieved a 95\% accuracy rate, underscoring its inherent strength in handling multimodal data, including chart data akin to time series image classification. Furthermore, it exhibits excellent classification performance on small samples such as T2CEP, with an $\rm F_1-$score of 0.94, which further validates the sensitivity of image-based models to small sample sizes. The LALM's innovative approach of converting time-series data into audio signals for classification led to a commendable accuracy rate of 93\%, highlighting the potential of audio-based analysis in scientific research. These results collectively emphasize the effectiveness and versatility of the StarWhisper LC series in leveraging transfer learning and LLM-based model for high-accuracy classification tasks in the realm of astrophysics.

\subsubsection*{Catalogs} \label{subsec:catalog}

Key factors affecting light curve classification include period, sampling rate, and morphological features. Therefore, we conducted an analysis and discussion on the importance of different phases and sampling rates for the Conv1D+LSTM model, which performed best among the models based on time series.

\begin{figure}[htbp]
    \centering
    \begin{subfigure}[b]{0.45\textwidth}
        \includegraphics[width=\textwidth]{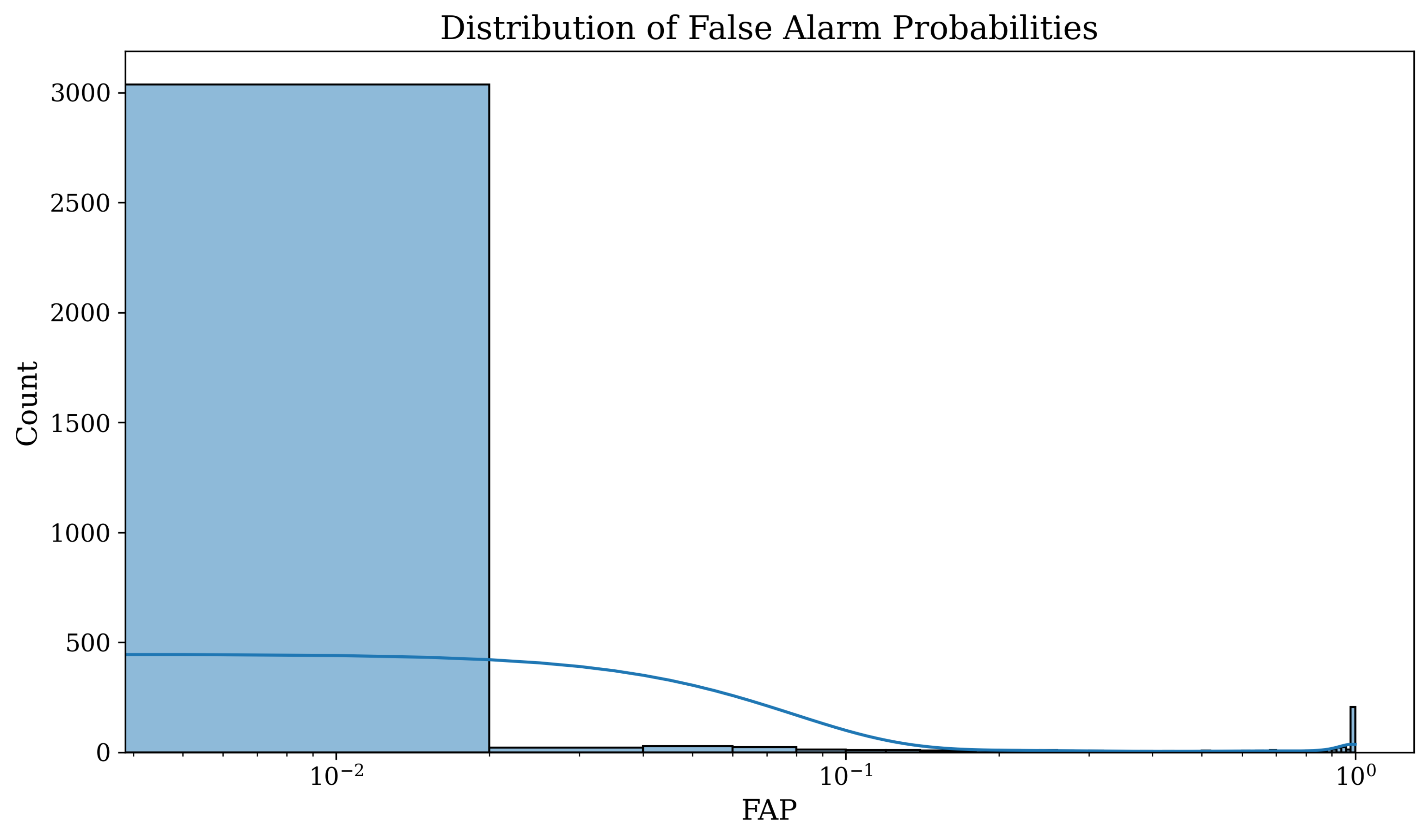}
        \caption{}
        \label{fig:fap_distribution}
    \end{subfigure}
    \hfill
    \begin{subfigure}[b]{0.45\textwidth}
        \includegraphics[width=\textwidth]{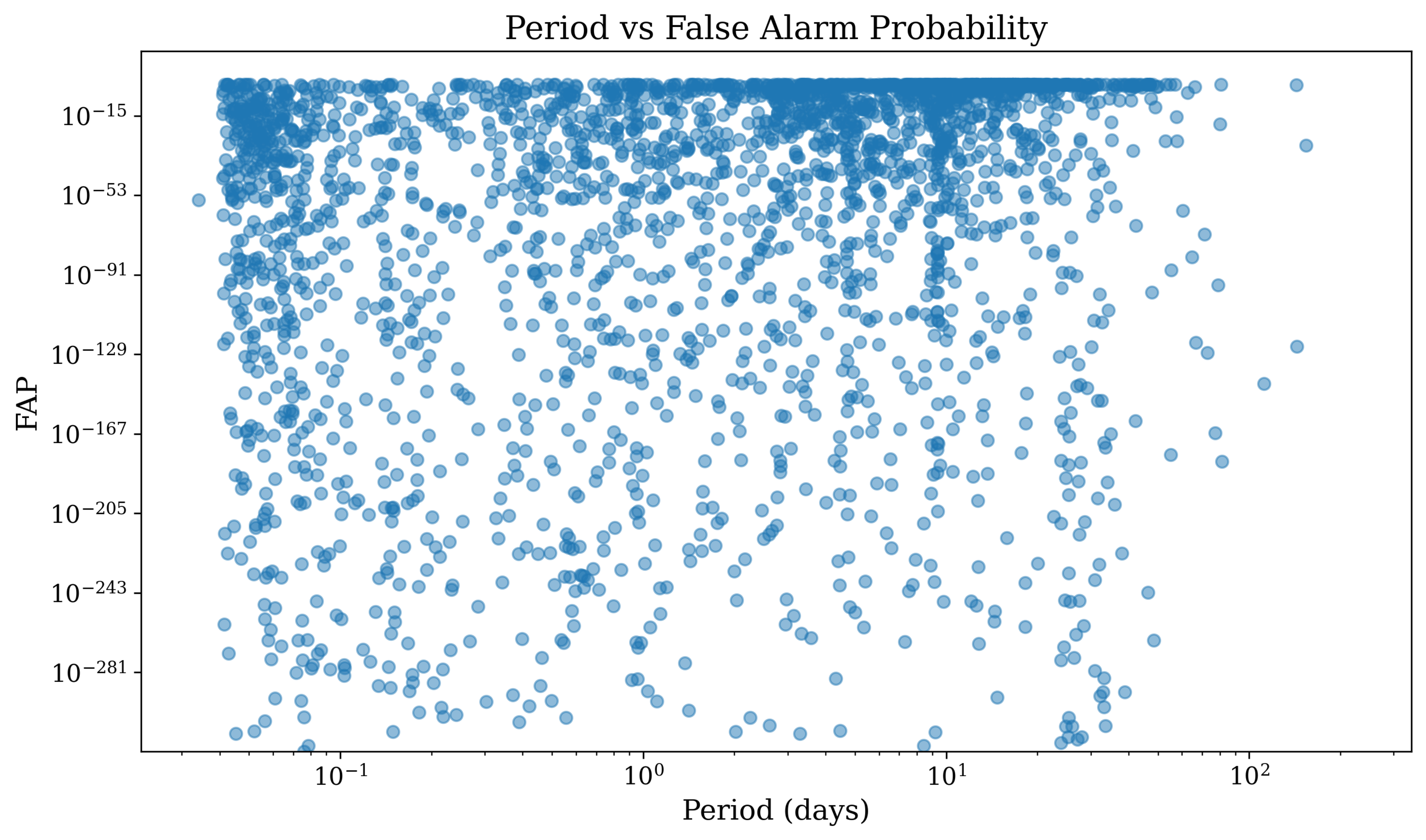}
        \caption{}
        \label{fig:period_vs_fap}
    \end{subfigure}
    \caption{Quantifying the probability of observing
a signal of equal or greater power in random noise data. (a) Distribution of False Alarm Probabilities. (b) Period vs False Alarm Probability.}
    \label{fig:combined}
\end{figure}

\emph{Phase importance}
The FAP, computed using the Lomb-Scargle algorithm, quantifies the probability of observing a signal of equal or greater power in random noise data. Our analysis shows that 81.35\% of the detected periods have FAP values below 0.001, indicating high statistical significance. Fig. \ref{fig:combined}(a) displays the distribution of FAP values, highlighting the concentration of periods with very low FAP. Fig. \ref{fig:combined}(b) shows the relationship between periods and their corresponding FAP values, offering insights into the reliability of detections across different timescales. Primarily attributable to the high-quality data from Kepler and K2 are the exceptionally low FAP values. These missions offer precise photometry, long observation baselines, and continuous monitoring, considerably boosting our ability to detect periodic signals accurately. With these statistically significant periods as our foundation, we proceed to conduct a detailed phase importance analysis to uncover critical phase-dependent features of variable stars.

\begin{figure*}[ht!]
\centering
\includegraphics[width=0.9\textwidth]{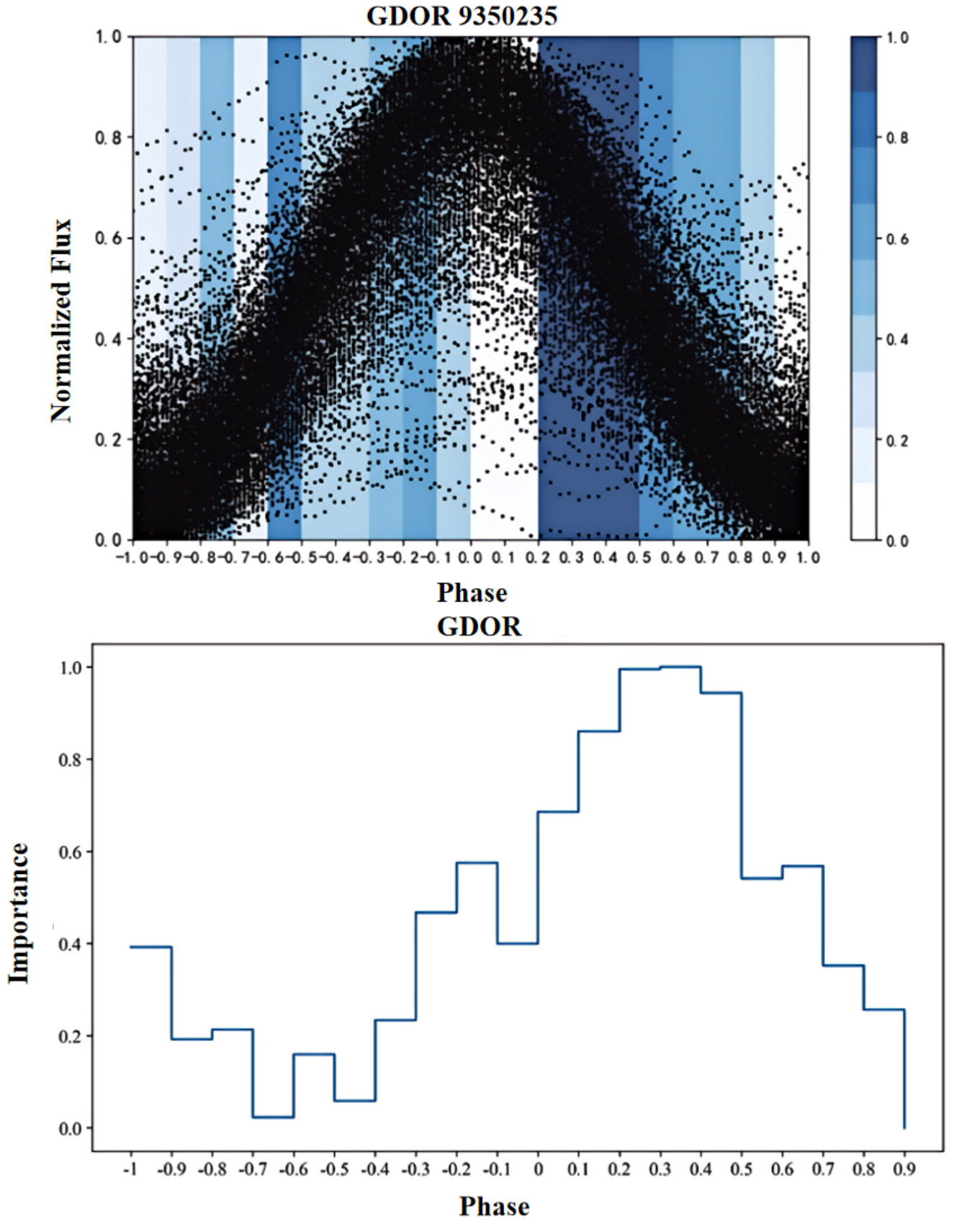}
\caption{Importance-phase diagram for a GDOR. The upper panel shows the light curve of a specific object with zero phase aligned to the peak, overlaid with a heatmap illustrating phase importance. The lower panel displays the importance-phase relationship, with the x-axis representing phases from -1 to 1 (in 0.1 increments) and the y-axis showing relative importance. Masked phase intervals from -1 to 1, with a step size of 0.1. It is represented by removing the corresponding data points in each 0.1 bin.}
\label{fig:1}
\end{figure*}

\begin{figure*}[ht!]
\centering
\includegraphics[width=0.9\textwidth]{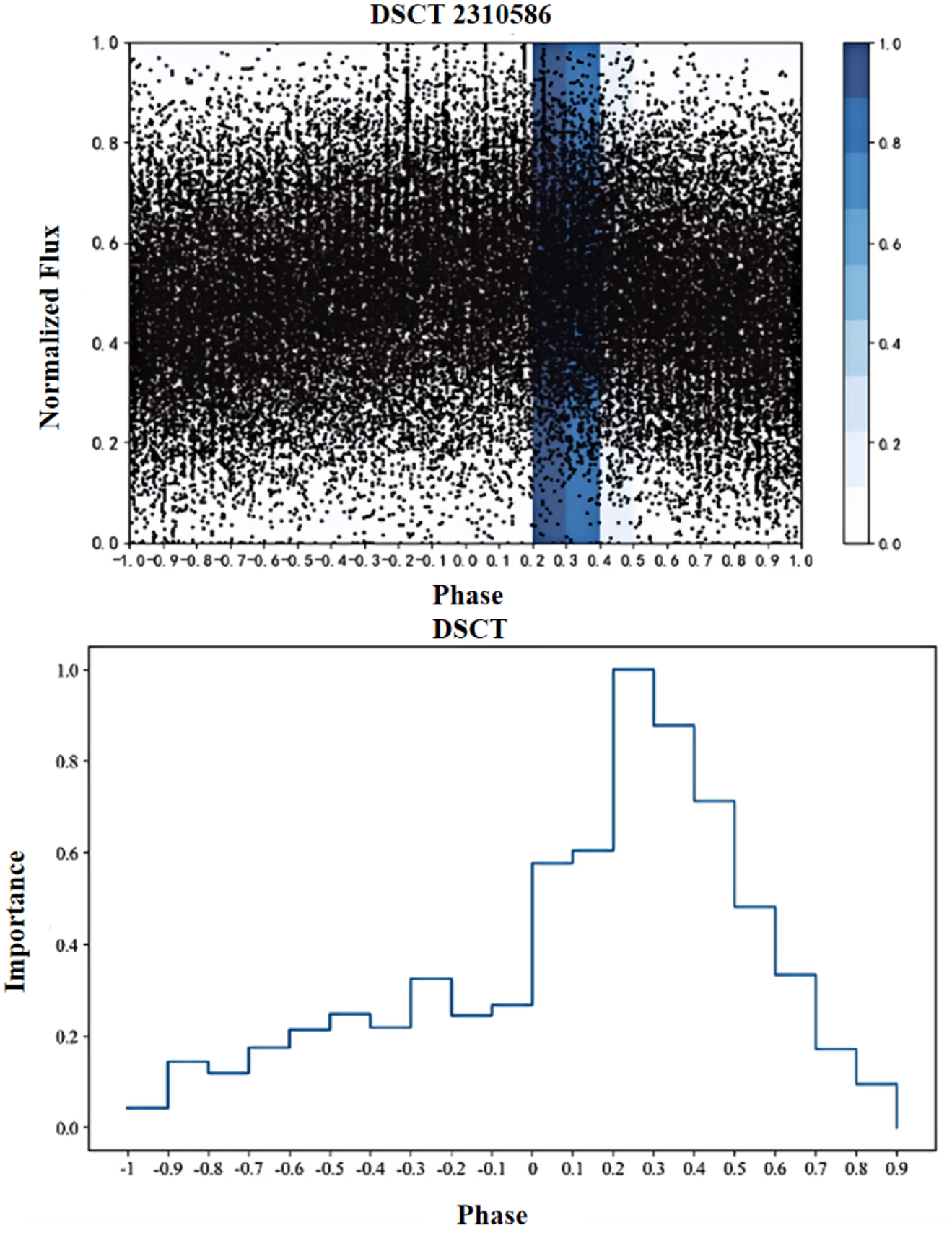}
\caption{Importance-phase diagram for a DSCT. The upper panel shows the light curve of a specific object with zero phase aligned to the peak, overlaid with a heatmap illustrating phase importance. The lower panel displays the importance-phase relationship, with the x-axis representing phases from -1 to 1 (in 0.1 increments) and the y-axis showing relative importance. Masked phase intervals from -1 to 1, with a step size of 0.1. It is represented by removing the corresponding data points in each 0.1 bin.}
\label{fig:2}
\end{figure*}

\begin{figure*}[ht!]
\centering
\includegraphics[width=0.9\textwidth]{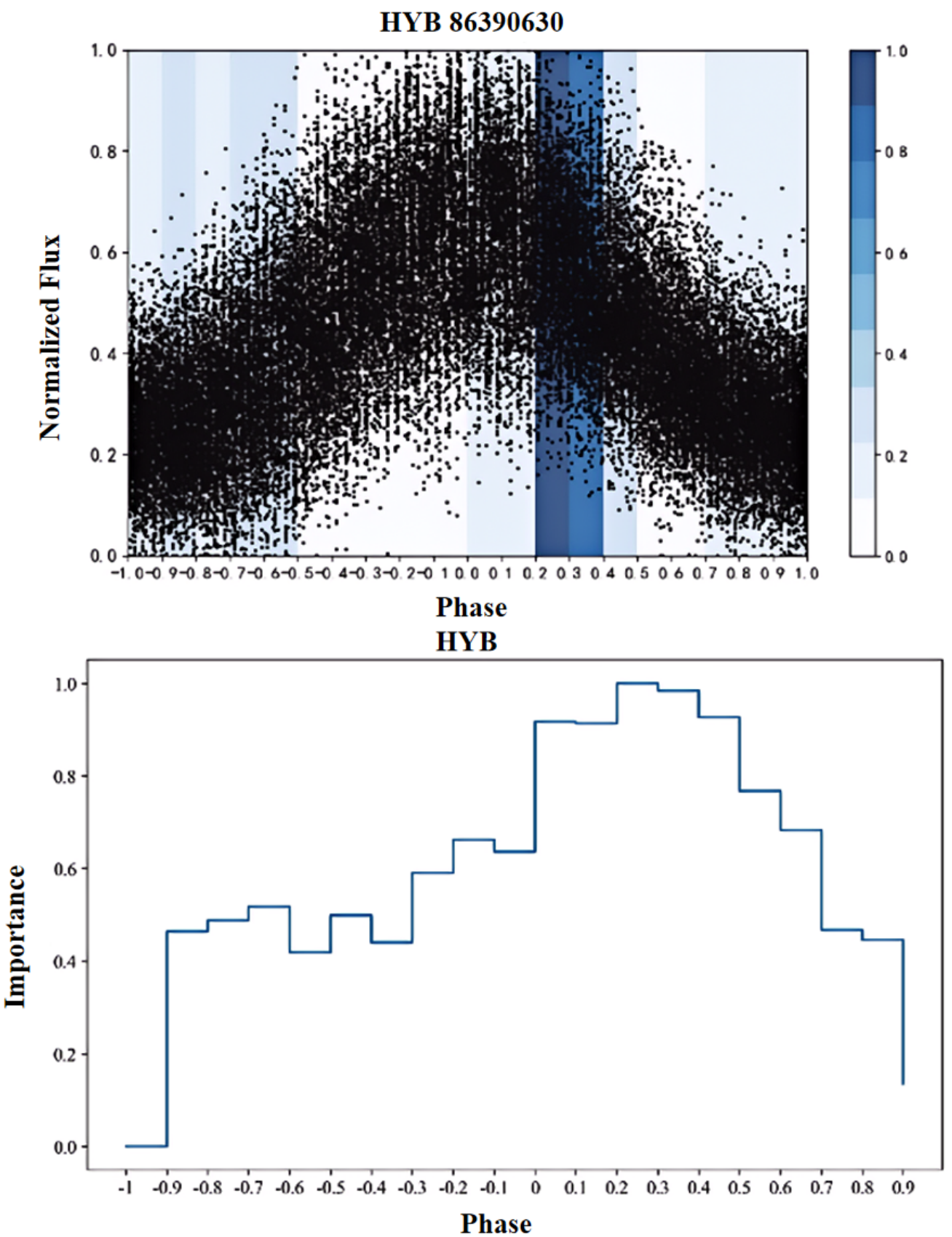}
\caption{Importance-phase diagram for an HYB. The upper panel shows the light curve of a specific object with zero phase aligned to the peak, overlaid with a heatmap illustrating phase importance. The lower panel displays the importance-phase relationship, with the x-axis representing phases from -1 to 1 (in 0.1 increments) and the y-axis showing relative importance. Masked phase intervals from -1 to 1, with a step size of 0.1. It is represented by removing the corresponding data points in each 0.1 bin.}
\label{fig:3}
\end{figure*}

\begin{figure*}[ht!]
\centering
\includegraphics[width=0.9\textwidth]{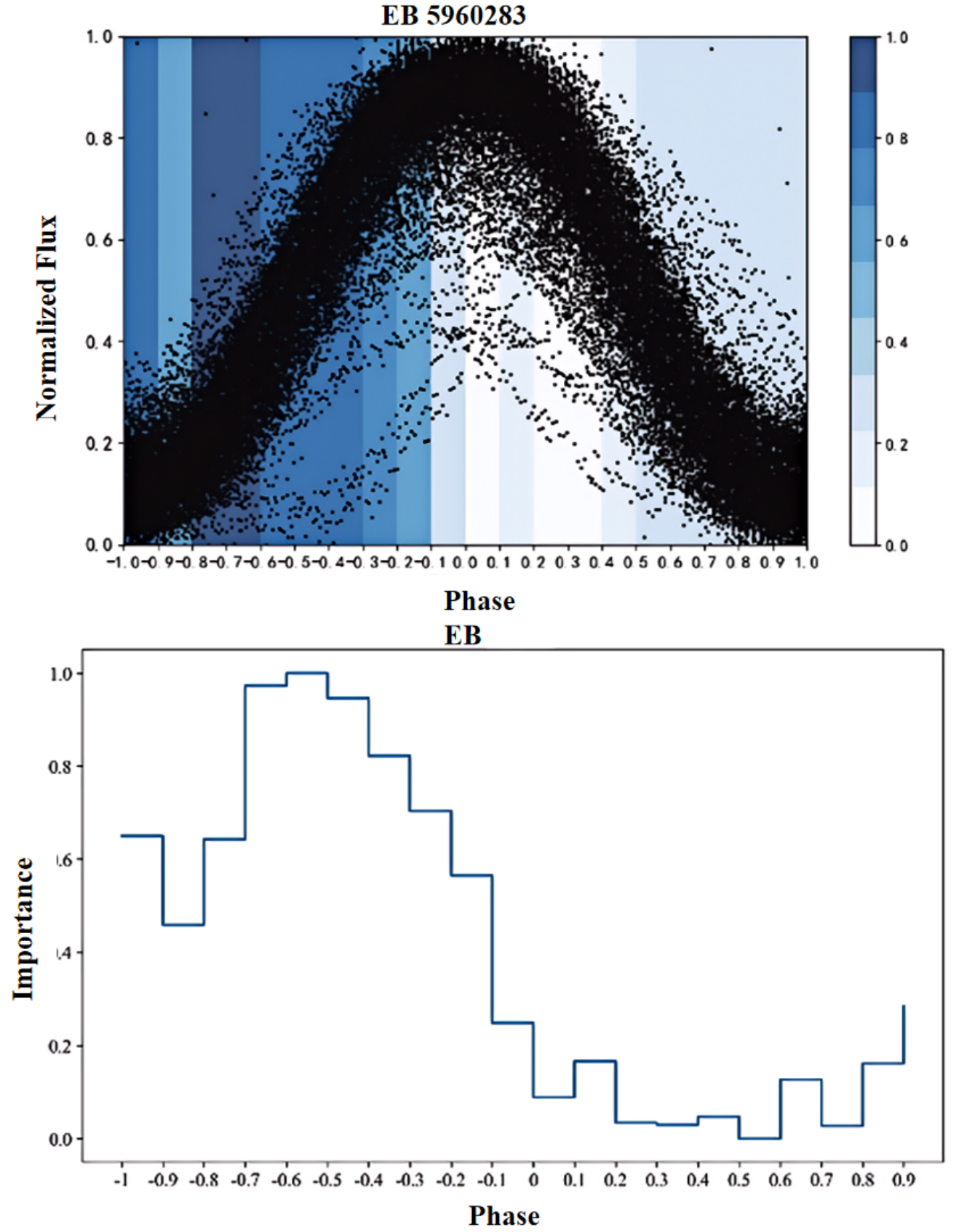}
\caption{Importance-phase diagram for an EB. The upper panel shows the light curve of a specific object with zero phase aligned to the peak, overlaid with a heatmap illustrating phase importance. The lower panel displays the importance-phase relationship, with the x-axis representing phases from -1 to 1 (in 0.1 increments) and the y-axis showing relative importance. Masked phase intervals from -1 to 1, with a step size of 0.1. It is represented by removing the corresponding data points in each 0.1 bin.}
\label{fig:4}
\end{figure*}

\begin{figure*}[ht!]
\centering
\includegraphics[width=0.9\textwidth]{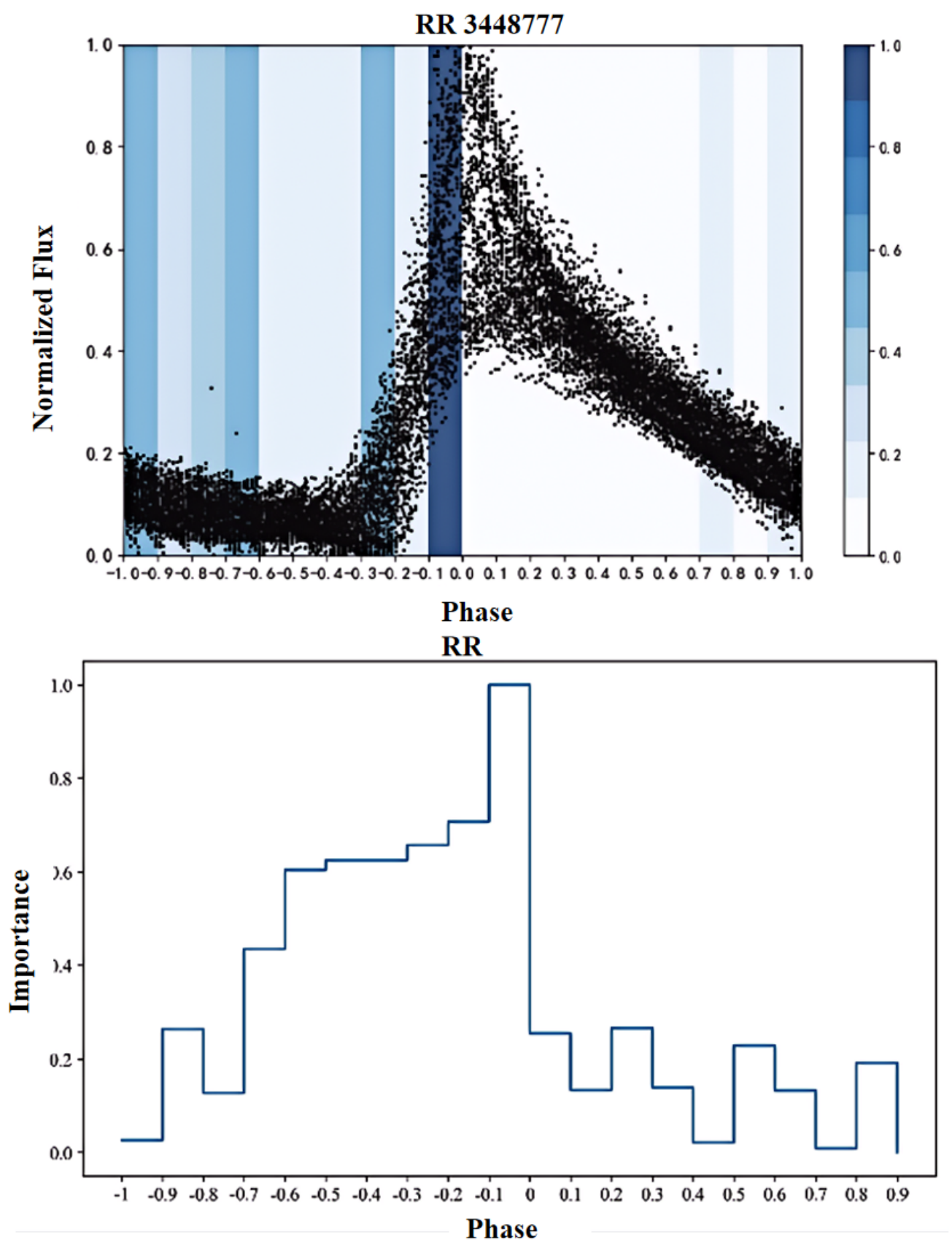}
\caption{Importance-phase diagram for an RR. The upper panel shows the light curve of a specific object with zero phase aligned to the peak, overlaid with a heatmap illustrating phase importance. The lower panel displays the importance-phase relationship, with the x-axis representing phases from -1 to 1 (in 0.1 increments) and the y-axis showing relative importance. Masked phase intervals from -1 to 1, with a step size of 0.1. It is represented by removing the corresponding data points in each 0.1 bin. }
\label{fig:5}
\end{figure*}

By aligning the zero phase with the peak of the light curve according to the period, we effectively restricted the model's access to specific segments. We systematically deleted phase intervals ranging from -1 to 1, in 0.1 increments, essentially masking each bin as shown in Figs. \ref{fig:1} to \ref{fig:5}. This approach allowed us to evaluate how removing each segment impacted the Conv1D + BiLSTM model's performance, effectively simulating the real-world scenario in time-domain astronomy where certain phase observations may be missing or discontinuous.

Table \ref{tab:phase} lists the importance values with an interval of 0.1 phase. Figs. \ref{fig:1}, \ref{fig:2} and \ref{fig:3} reveal that the key features for the GDOR, DSCT, and HYB, are primarily located in the phase interval following the peak flux. For the EB and RR (shown in Figs. \ref{fig:4} and \ref{fig:5}), the main concentration is detected in the phase interval during which the flux returns to its peak. According to the heatmap distribution, significant or dark intervals predominantly occur in the first half of the EB and RR. In contrast, DSCTs and HYBs show 2 distinct phase intervals localized within them. We discuss the time saving associated with this and present a catalog in the Discussion section. 

\begin{table*}[!t]
\small
    \centering
    \caption{Phase Importance Catalog} 
    \label{tab:phase} 
    \begin{tabular}{lcccccccccc}
    \hline
    \hline
    & \multicolumn{10}{c}{\textbf{Focused phase}}\\
        \textbf{Star} & \textbf{0} & \textbf{0.1} & \textbf{0.2} & \textbf{0.3} & \textbf{0.4} & \textbf{0.5} & \textbf{0.6} & \textbf{0.7} & \textbf{0.8} & \textbf{0.9} \\ \hline
        DSCT\_001026294 & 0.9996  & 0.9934  & 1.0000  & 1.0000  & 0.9995  & 0.9986  & 0.9986  & 0.9692  & 0.6977  & 0.0000  \\
        DSCT\_001162150 & 0.1585  & 0.1310  & 1.0000  & 0.7921  & 0.3214  & 0.1172  & 0.1245  & 0.0323  & 0.0184  & 0.0000  \\
        DSCT\_001163943 & 0.3598  & 0.1016  & 1.0000  & 0.4261  & 0.0467  & 0.1701  & 0.0711  & 0.0031  & 0.0393  & 0.0188  \\
        DSCT\_001294670 & 0.3018  & 0.3963  & 1.0000  & 0.6999  & 0.0631  & 0.0211  & 0.0327  & 0.0001  & 0.0034  & 0.0004  \\
        DSCT\_001430590 & 0.1579  & 0.2669  & 1.0000  & 0.5738  & 0.4043  & 0.0964  & 0.0573  & 0.0266  & 0.0102  & 0.0041  \\
        DSCT\_001434660 & 0.0923  & 0.1533  & 1.0000  & 0.6423  & 0.6384  & 0.1624  & 0.0591  & 0.0247  & 0.0037  & 0.0142  \\
        DSCT\_001570023 & 0.0688  & 0.1608  & 1.0000  & 0.5841  & 0.4261  & 0.0822  & 0.0513  & 0.0040  & 0.0000  & 0.0027  \\
        DSCT\_001571717 & 0.9996  & 0.9994  & 1.0000  & 0.9999  & 1.0000  & 0.9997  & 1.0000  & 0.9942  & 0.9730  & 0.0000  \\
        DSCT\_001572768 & 0.0000  & 0.9147  & 0.9228  & 0.9773  & 0.9816  & 0.8840  & 0.8007  & 0.8885  & 0.4628  & 0.4773  \\
        DSCT\_001575977 & 0.1454  & 0.6365  & 1.0000  & 0.8420  & 0.2347  & 0.3248  & 0.3394  & 0.0303  & 0.0057  & 0.0241 \\ \hline
    \end{tabular}
            \begin{tablenotes}
            \small
            \item Phase-importance catalog illustrating the relationship between importance and phase, with the first 10 rows shown.
        \end{tablenotes}
\end{table*}

\emph{Sampling}

\begin{figure*}[ht!]
\centering
\includegraphics[width=\textwidth]{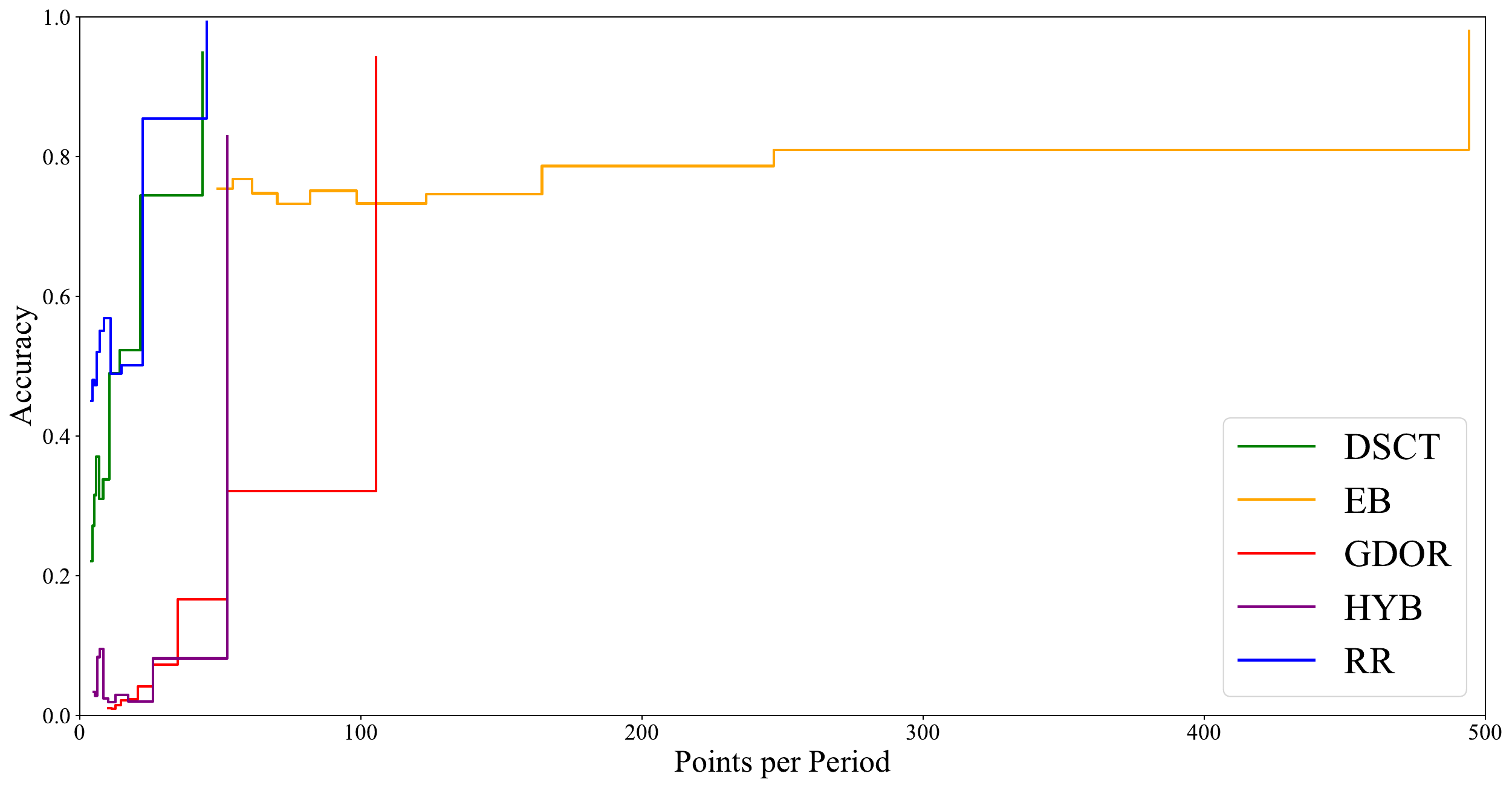}
\caption{The influence of each period sampling point on model accuracy.
\label{fig:output}}
\end{figure*}

The observation schedule is considerably influenced by sampling, which itself is constrained by factors such as telescope aperture, CCD read-out speed, and the scientific objectives of the survey. To comprehensively understand the impact of sampling on variable star classification, we conducted a detailed sampling-importance relationship analysis for variables with a single prime frequency. By systematically adjusting the number of sampling points per period, we simulated various observational conditions and thoroughly investigated their effect on classification accuracy. Fig. \ref{fig:output} clearly illustrates this complex relationship. 

Our analysis revealed distinct response patterns of different variable star types to changes in sampling rates. EBs demonstrated remarkable robustness to sampling rate variations, maintaining an accuracy above 75$\%$ even when sampling points were reduced to 1/10 of the original. This finding suggests that the light curve features of EBs can be effectively identified even at lower temporal resolutions, possibly due to their unique periodic luminosity variation patterns. For RR and DSCT stars, halving the sampling rate still maintained an accuracy of around 80$\%$. This stability likely stems from the strong periodicity and characteristic light curve shapes of these stars, preserving their key features even at lower sampling rates. 

We observed a general phenomenon where the prediction model's accuracy considerably decreases beyond a certain sampling rate threshold. This critical point varies among star types, reflecting differences in their sensitivity to temporal resolution. For instance, DSCT stars showed a sharp decline in accuracy when the sampling rate was reduced to about 1/4 of the original, while EBs only exhibited a noticeable decrease at around 1/10. Through analyzing classification accuracies at different sampling rates, we found a clear correlation between sampling rate and phase importance. This not only validates our methodology but also provides new insights into understanding key phases in variable star light curves. 

In addition, at specific lower sampling intervals, some variable types exhibited an anomalous decrease in accuracy with increasing sampling. This phenomenon was particularly noticeable in RR stars when sampling points were reduced to about 1/8 of the original. We hypothesize that within this sampling interval, specific instrumental noise or subtle light curve features may be erroneously amplified, leading to a temporary decrease in accuracy. This reveals a critical sampling threshold where feature capture and noise influence reach a delicate balance point. The sensitivity to sampling rate changes varied considerably among different types of variable stars. For example, EBs showed high robustness at low sampling rates, while DSCT stars were more sensitive to sampling rate reductions. This disparity likely reflects differences in the intrinsic physical properties and light variation mechanisms of different variable star types.

Table \ref{tab:sampling} presents detailed classification accuracies for various variable star types at different sampling rates. These results not only deepen our understanding of sampling rate effects but also provide crucial guidance for future variable star observation strategies. For instance, differentiated sampling strategies can be developed for different types of variable stars, optimizing observational resources while ensuring classification accuracy. 
 
\begin{table*}[!t]
    \centering
    \caption{Sampling Catalog} 
    \label{tab:sampling} 
    \begin{tabular}{lcccccccccc}
    \hline
    \hline
        \textbf{Star/Sampling Rate} & \textbf{0.02d} & \textbf{0.04d} & \textbf{0.06d} & \textbf{0.08d} & \textbf{0.1d} & \textbf{0.12d} & \textbf{0.14d} & \textbf{0.16d} & \textbf{0.18d} & \textbf{0.2d} \\ \hline
        DSCT\_001026294 & 0.9288  & 0.4412  & 0.0002  & 0.0003  & 0.0000  & 0.0000  & 0.0434  & 0.0847  & 0.0998  & 0.0141  \\
        DSCT\_001162150 & 0.9817  & 0.9598  & 0.1599  & 0.6392  & 0.3082  & 0.1531  & 0.4663  & 0.2866  & 0.1713  & 0.1649  \\
        DSCT\_001163943 & 0.9891  & 0.7806  & 0.6678  & 0.5106  & 0.5077  & 0.6081  & 0.6654  & 0.6412  & 0.5840  & 0.4549  \\
        DSCT\_001294670 & 0.9985  & 0.9980  & 0.9521  & 0.7222  & 0.3226  & 0.6149  & 0.5517  & 0.5106  & 0.5157  & 0.2001  \\
        DSCT\_001430590 & 0.9935  & 0.9918  & 0.2952  & 0.6864  & 0.6574  & 0.2160  & 0.1984  & 0.5479  & 0.2953  & 0.3322  \\
        DSCT\_001434660 & 0.9746  & 0.9851  & 0.8613  & 0.3703  & 0.4243  & 0.3810  & 0.5342  & 0.2130  & 0.3090  & 0.2803  \\
        DSCT\_001570023 & 0.9944  & 0.9897  & 0.8082  & 0.8866  & 0.4413  & 0.2945  & 0.5183  & 0.4653  & 0.2904  & 0.1887  \\
        DSCT\_001571717 & 0.9656  & 0.0958  & 0.0018  & 0.0159  & 0.0000  & 0.0001  & 0.3438  & 0.0925  & 0.0448  & 0.0198  \\
        DSCT\_001572768 & 0.9253  & 0.3443  & 0.1441  & 0.1154  & 0.1095  & 0.0479  & 0.0768  & 0.0454  & 0.1152  & 0.1194  \\
        DSCT\_001575977 & 0.9839  & 0.7853  & 0.6452  & 0.3186  & 0.3033  & 0.3250  & 0.2376  & 0.3029  & 0.2770  & 0.2040 \\ \hline
    \end{tabular}
            \begin{tablenotes}
            \small
            \item Sampling catalog illustrating the relationship between accuracy and sampling rate, with the first 10 rows shown.
        \end{tablenotes}
\end{table*}

\subsection*{Discussion} \label{sec:discussion}

\subsubsection*{Observation time saving}

By correlating classification accuracy across variables with phase importance, we sequentially eliminate multiple phase intervals in ascending order of significance. This process help us estimate the maximum observation time that can be saved for each star type. It is suggested that, on average, a 14$\%$ reduction in observation time can be achieved across different variables, provided the accuracy variation remains within 10$\%$ over a given observation period. Specifically, RR and EB stars can save an average of 44$\%$ and 29$\%$ of observation time, respectively.

In addition to analyzing classification accuracy across varying sampling rates, we investigated the relationship between accuracy and the number of sampling points within each period to assess the potential for reducing observation time. Our findings reveal that accuracy variations within a single observation period generally remain under 10$\%$, indicating important opportunities for optimizing data collection strategies. On average, we found that a 21$\%$ reduction in the number of sampling points is possible without substantial loss in classification accuracy. Notably, for EB stars, an average reduction of 54$\%$ in sampling points is achievable while maintaining robust classification performance. These results, presented in Table \ref{tab:Save}, have profound implications for observational astronomy.

Our proposed RNN-based models demonstrate robust performance without requiring extensive image preprocessing, further enhancing their practical utility. The use of automated deep learning techniques facilitates efficient hyperparameter optimization, contributing to the models' effectiveness. Additionally, we evaluated the impact of masked or unobserved points within the complete phase on classification accuracy. Remarkably, EB stars maintain an accuracy around 0.75 even with sampling points reduced to 10\%, while DSCT stars and RR variables sustain accuracies of approximately 0.75 and 0.85 respectively when sampling is halved. These findings underscore the resilience of our method to incomplete or sparse data, a common challenge in astronomical observations due to various constraints such as weather conditions or instrumental limitations. The robustness of our approach to reduced sampling makes it particularly suitable for efficient, and potentially real-time, astronomical time-series recognition scenarios. This capability has wide-ranging applications, from exoplanet detection \cite{salinasDistinguishingPlanetaryTransit2023} to the identification of transient events \cite{muthukrishnaRealTimeDetectionAnomalies2022}. 

In the context of large-scale surveys, our method could considerably enhance the speed of initial classification, allowing for rapid follow-up decisions on interesting targets. Moreover, the ability to maintain high classification accuracy with reduced data points could be instrumental in managing the data deluge expected from next-generation telescopes and surveys. By enabling efficient processing of vast datasets with optimized sampling, our approach could help astronomers navigate the challenges of big data in astronomy while considerably increasing scientific output. 

\subsubsection*{Learning from imbalanced data} \label{subsec:T2CEP}

Imbalanced samples are a common occurrence in astronomical data. In this study, we have applied transfer learning models that were previously trained on a considerable volume of image data to augment the recognition capabilities for these less prevalent variable stars. Notably, the Swin Transformer and MLLM yield moderately good results.

Additionally, we find that the self-attention mechanism can considerably enhance the ability to recognize small samples. We will further consider applying data augmentation, resampling, pretraining, and other methods in deep learning models based on self-attention, in order to improve their performance on tasks involving imbalanced data, such as data on variable stars collected by the Transiting Exoplanet Survey Satellite (TESS).

\subsubsection*{Time series as language}

\begin{figure*}[ht!]
\centering
\includegraphics[width=\textwidth]{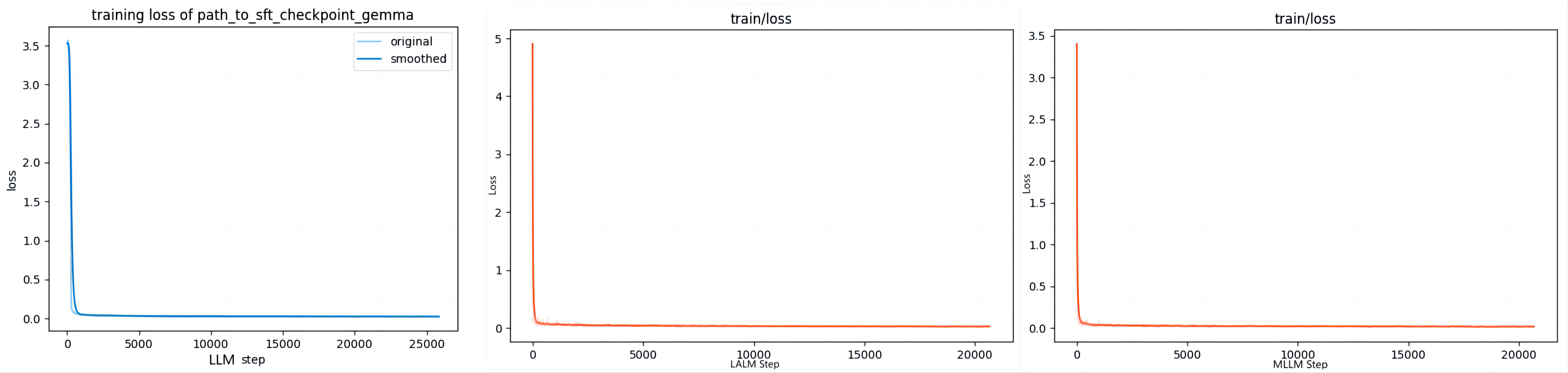}
\caption{Loss curves of LLM-based methods.
\label{fig:loss}}
\end{figure*}

Our research highlights the potential of leveraging the emergent capabilities of LLM-based models for the processing of light curves, a task that requires rapid convergence as shown in Fig. \ref{fig:loss}, resistance to overfitting, and minimal susceptibility to data quality variations \cite{liuSiTianProject2021}. By integrating these models with additional capability modules, such as visual and audio encoding, we aim to enhance their performance and applicability in the domain of astronomy. This approach not only introduces a novel analytical method for astronomical data but also explore the development of multimodal models that can process a variety of astronomical inputs.

The optimization of LLMs for inferential tasks, coupled with their capacity for parallel and rapid data processing \cite{kwon2023efficient}, underscores their utility in handling the vast and complex datasets encountered in astronomy. There is a promising prospect of training specialized astronomical encoding modules that build upon the robust foundation of LLMs. Such modules could be tailored to interpret and analyze astronomical phenomena with a high degree of accuracy and efficiency.

In the future, we will focus on refining the models by adjusting parameters such as data volume, sampling points, precision, and temporal length. This will enable us to delve deeper into the feature extraction thresholds of the models and further enhance their capabilities for astronomy-specific tasks. Moreover, the potential for applying this methodology to other time series tasks is considerable, offering a versatile path for developing multitask applications that operate on a multimodal basis.

\subsubsection*{SiTian project} \label{subsec:sitian}

The SiTian prototype, introduced in \cite{liuSiTianProject2021}, iplans to share its internal data. The data will contain information from 3 bands, and we plan to construct its light curves using numerical simulations and evaluate their classifications using deep learning techniques. Given the large volume of data, striking a balance between prediction time and accuracy will be essential. 

Highly accurate models, particularly models that are sensitive to smaller samples, would considerably enhance SiTian's capabilities in identifying variable stars. The catalogs we have developed related to phase importance and sampling provide insights into optimal observation intervals and sampling rates during monitoring. This allows for a calculated tradeoff between model accuracy and the time costs associated with training and prediction, ensuring efficient and effective astronomical analyses.

\subsubsection*{Related work}
\cite{Hinners_2018editor} performed a study to determine the most effective features from the feature analysis for time series (FATS) library for light curve classification and stellar parameter regression tasks. By employing ridge regression, they discovered that the optimal outcomes were obtained using the mean, standard deviation, skewness, kurtosis, and the period of the light curve. Among these, the features that had the greatest influence included the frequency $n$ harmonics amplitude $i$, frequency $n$ harmonics relative phase $i$, maximum slope, linear trend, and log likelihood ratio. These features encapsulate the essential characteristics of the light curves, whereas other features necessitate further processing to reflect more complex aspects of the light curves. In contrast to \cite{Hinners_2018editor}, our focus lies primarily on selecting features directly from the raw light curve data, eliminating the need for feature extraction. As a result, the classification accuracies achieved in our method surpass those reported in \cite{Hinners_2018editor}, owing to the testing of novel methods that possess enhanced classification capabilities. 

\cite{Charnock17RNN} applied a bidirectional RNN architecture to the Supernovae Photometric Classification Challenge (SPCC) light curve data, successfully classifying supernovae into types I, II, and III with an accuracy of 90.4\%. In another study, \cite{malik21exo} utilized the LightGBM method on Kepler and TESS data to detect exoplanets through the transit method, achieving 98\% accuracy in exoplanet classification and 82\% recall in TESS data. Similar to our work in T2CEP, these studies evaluated their classification methods on datasets with biased light curve distributions. Our results demonstrate higher accuracy (99\% total accuracy and 83\% accuracy on biased classes) in these biased classes (see Table \ref{tab:accuracy}), suggesting that recent advancements in methodologies have considerably enhanced the capability to handle classification tasks involving biased classes.

\section*{Conclusions} \label{sec:sec4}
In this study, we have thoroughly investigated the application of deep learning and LLMs for classifying variable star light curves using data from the Kepler and K2 missions. Our focus encompassed 5 types of variable stars—DSCT, GDOR, RR Lyrae, Hybrid, and eclipsing binaries—and included performance testing on small sample sets of type II Cepheids. Through automated deep learning optimization, we achieved exceptional performance with both a Conv1D + BiLSTM architecture and the Swin Transformer, achieving accuracies of 94\% and 99\%, respectively. Notably, the Swin Transformer excelled in identifying the rare Type II Cepheids, which make up only 0.02\% of the dataset, with an impressive accuracy of 83\%.

We also introduced StarWhisper LightCurve, an innovative series comprising 3 LLM-based models: LLM, MLLM, and LALM. These models are fine-tuned through strategic prompt engineering and customized training methods to harness their emergent capabilities for astronomical data analysis. The StarWhisper LightCurve models exhibit high accuracies around 90\%, considerably reducing the need for explicit feature engineering and paving the way for streamlined parallel data processing and the advancement of multifaceted multimodal models in astronomy.

Additionally, our research underscores the critical influence of observational cadence and phase distribution on classification precision. We provide 2 detailed catalogs illustrating how phase and sampling intervals impact deep learning classification accuracy. The findings reveal that reductions of up to 14\% in observation duration and 21\% in sampling points can be achieved without compromising accuracy by more than 10\%.

\section*{Acknowledgments}
The research presented in this paper was generously funded by the National Programs on Key Research and Development Project, with specific contributions from grant numbers 2019YFA0405504 and 2019YFA0405000. Additional support came from the National Natural Science Foundation of China (NSFC) under grants NSFC-11988101, 11973054, and 11933004. We also received backing from the Strategic Priority Program of the Chinese Academy of Sciences, granted under XDB41000000. Special acknowledgment goes to the China Manned Space Project for their science research grant, denoted by NO.CMS-CSST-2021-B07. We also acknowledge the Science and Education Integration Funding of University of Chinese Academy of Sciences.

JFL extends gratitude for support received from the New Cornerstone Science Foundation, particularly via the NewCornerstone Investigator Program, and the honor of the XPLORER PRIZE. 

This research incorporates data sourced from the Kepler mission, with its funding being attributed to the NASA Science Mission Directorate. We sourced all data for this study from the Mikulsk Archive for Space Telescopes (MAST). The operation of STScI is overseen by the Association of Universities for Research in Astronomy, Inc., under the NASA contract NAS5-26555. The MAST's support for non-HST data comes through the NASA Office of Space Science, notably grant NNX09AF08G, and various other grants and contracts.

\section*{Supplementary Materials}
Figs. S1 to S10\\

\printbibliography
\thispagestyle{empty}
\setcounter{figure}{0}  
\renewcommand{\thefigure}{S\arabic{figure}}
\setcounter{equation}{0}
\renewcommand{\theequation}{S\arabic{equation}}  
\setcounter{table}{0}  
\renewcommand{\thetable}{S\arabic{table}}  
\setcounter{section}{0}  
\renewcommand{\thesection}{S\arabic{section}}
\setcounter{page}{1}

\begin{centering}

\vspace{1cm}
{\Large Supplementary Materials  for\\}
{\LARGE Deep Learning and Methods Based on Large Language Models Applied to Stellar Light Curve Classification \\}
\vspace{1cm}
{\large Yu-Yang Li, Yu Bai, Cunshi Wang, Mengwei Qu, Ziteng Lu, Roberto Soria, and Jifeng Liu\\}
\vspace{.5cm}
{\large Corresponding author: Cunshi Wang, wangcunshi@nao.cas.cn\\}
\end{centering}

\vspace{2cm}
\noindent \textbf{The PDF file includes:}\\
\indent Figs. S1 to S10\\

\newpage

\section*{Supplementary Figures}

\begin{figure*}[ht!]
\centering
\includegraphics[width=0.9\textwidth]{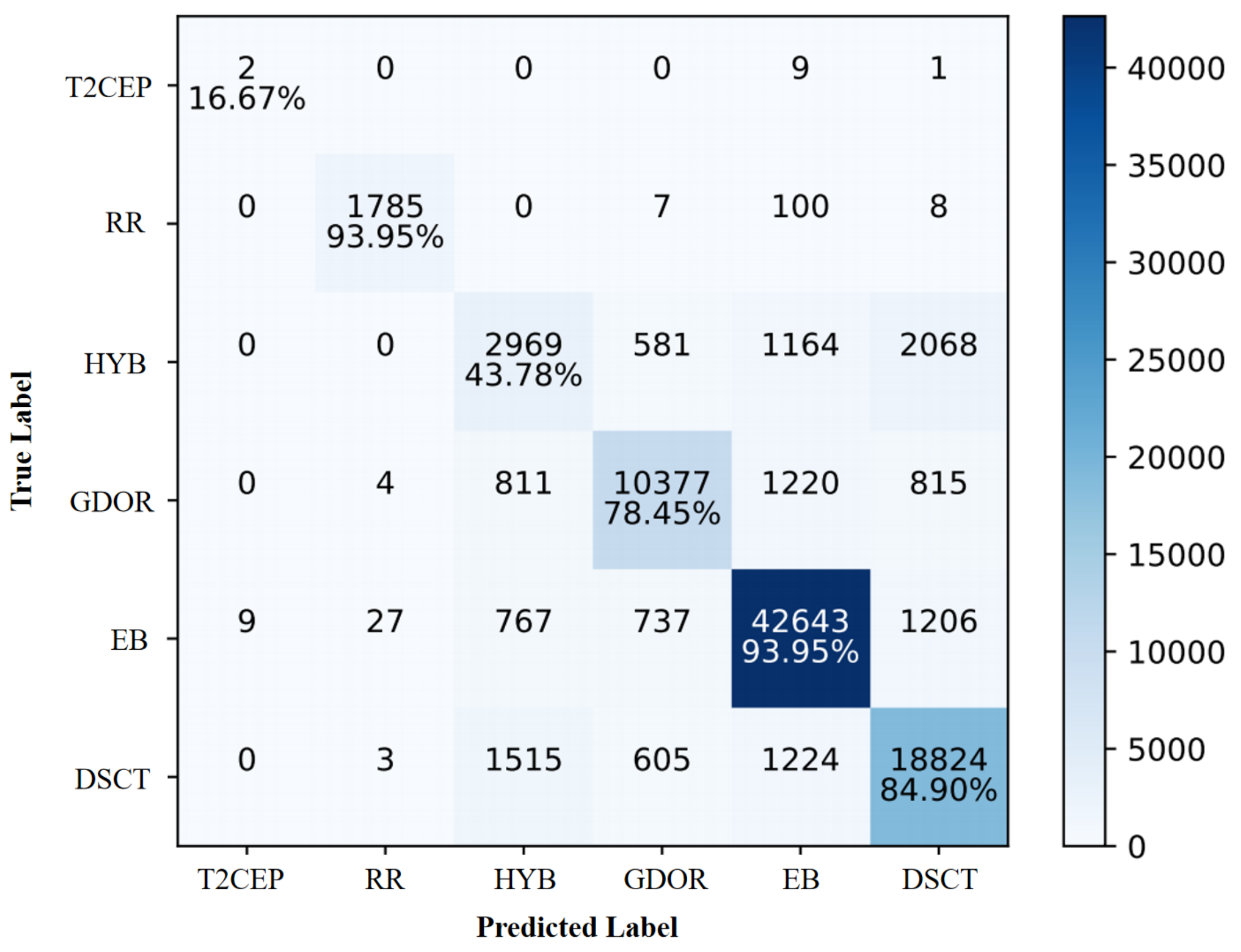}
\caption{The confusion matrix of the Conv1D + Transformer model is presented, including the T2CEP class. The diagonal elements represent the correct predictions, along with the percentages of correctly classified observations for each true class.
\label{fig:transformer}}
\end{figure*}

\begin{figure*}[ht!]
\centering
\includegraphics[width=0.9\textwidth]{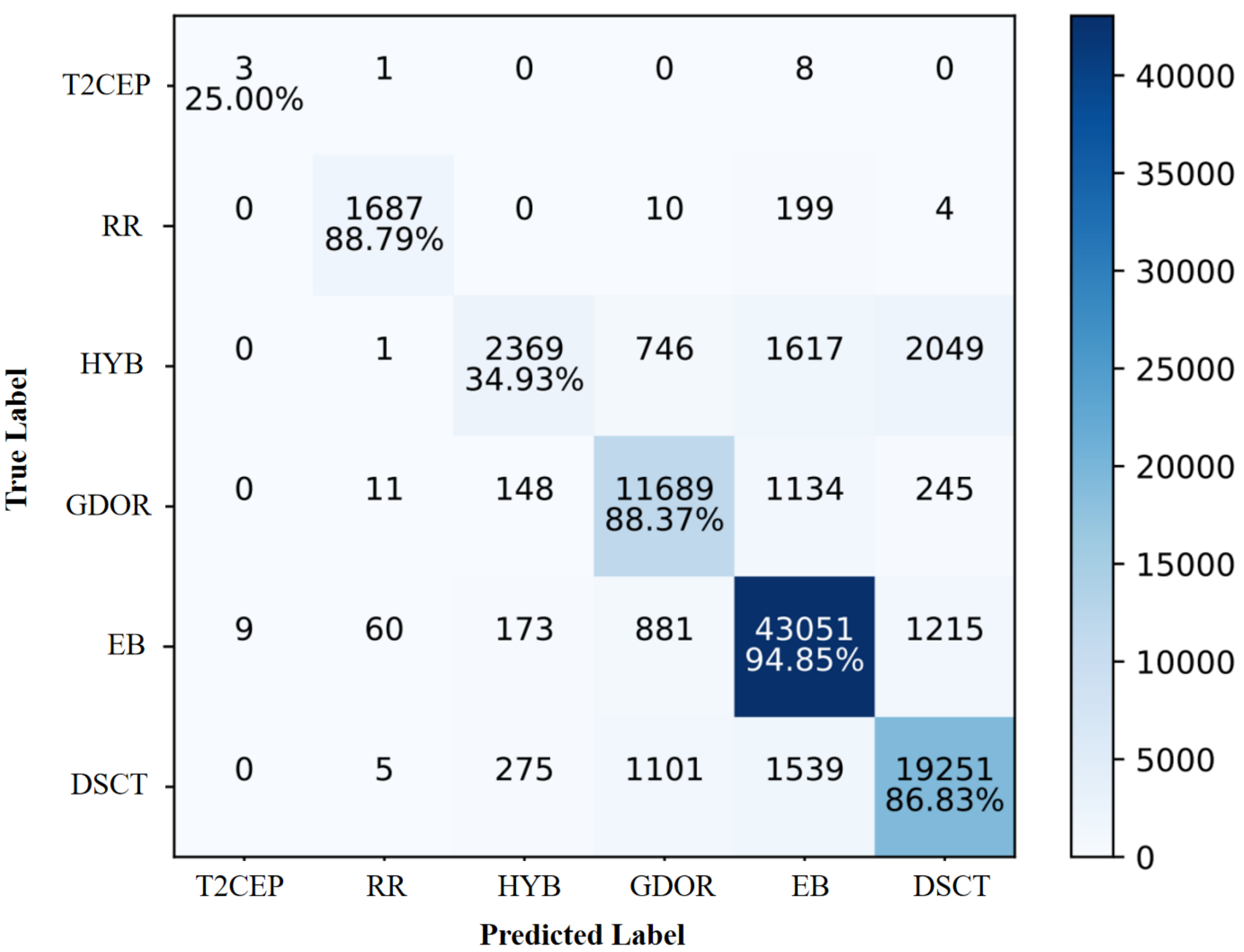}
\caption{The confusion matrix of the LightGBM model is presented, including the T2CEP class. The diagonal elements represent the correct predictions, along with the percentages of correctly classified observations for each true class.
\label{fig:lightgbm}}
\end{figure*}

\begin{figure*}[ht!]
\centering
\includegraphics[width=0.9\textwidth]{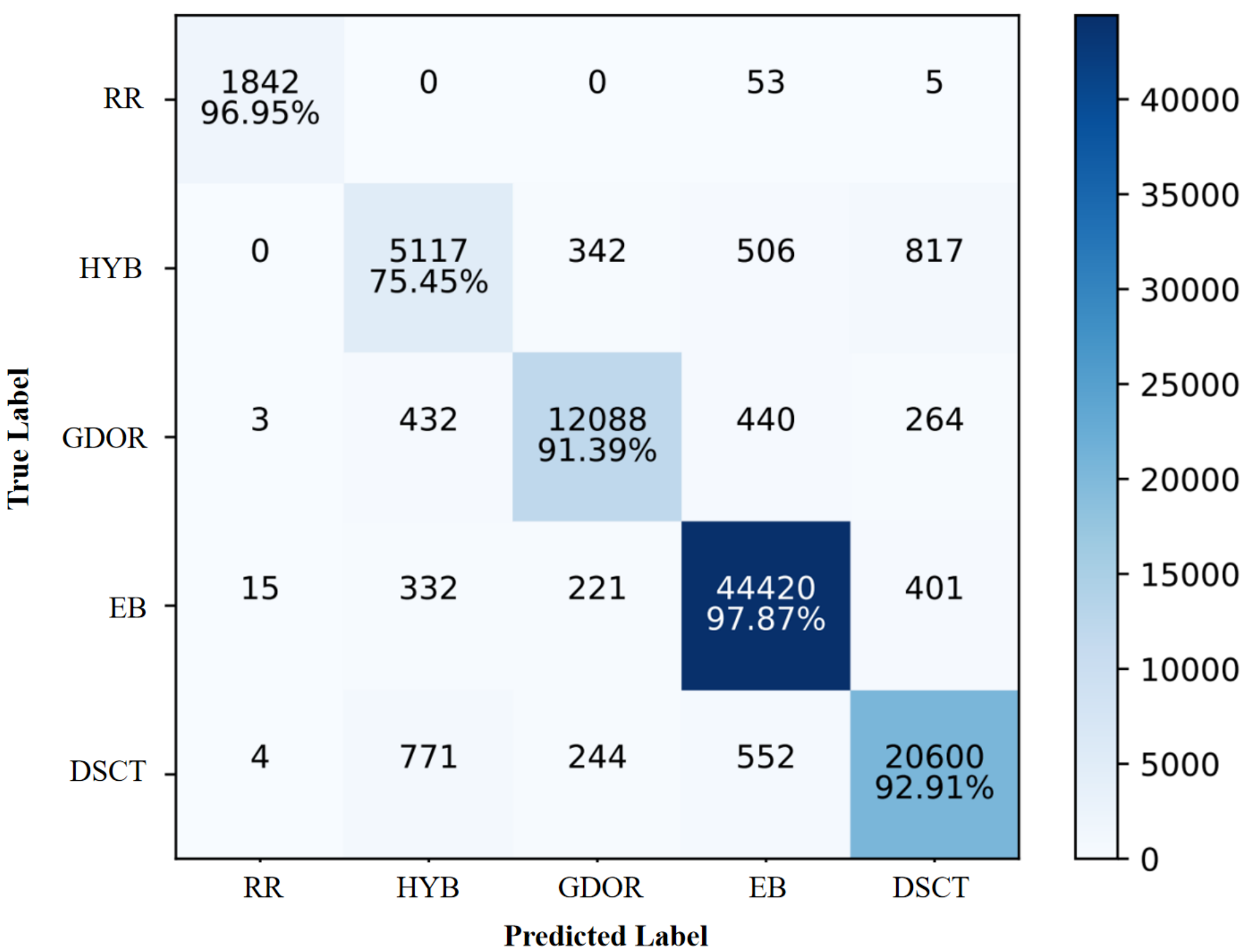}
\caption{The confusion matrix of the BiLSTM + Attention model. The diagonal elements represent the correct predictions, along with the percentages of correctly classified observations for each true class.
\label{fig:attention}}
\end{figure*}

\begin{figure*}[ht!]
\centering
\includegraphics[width=0.9\textwidth]{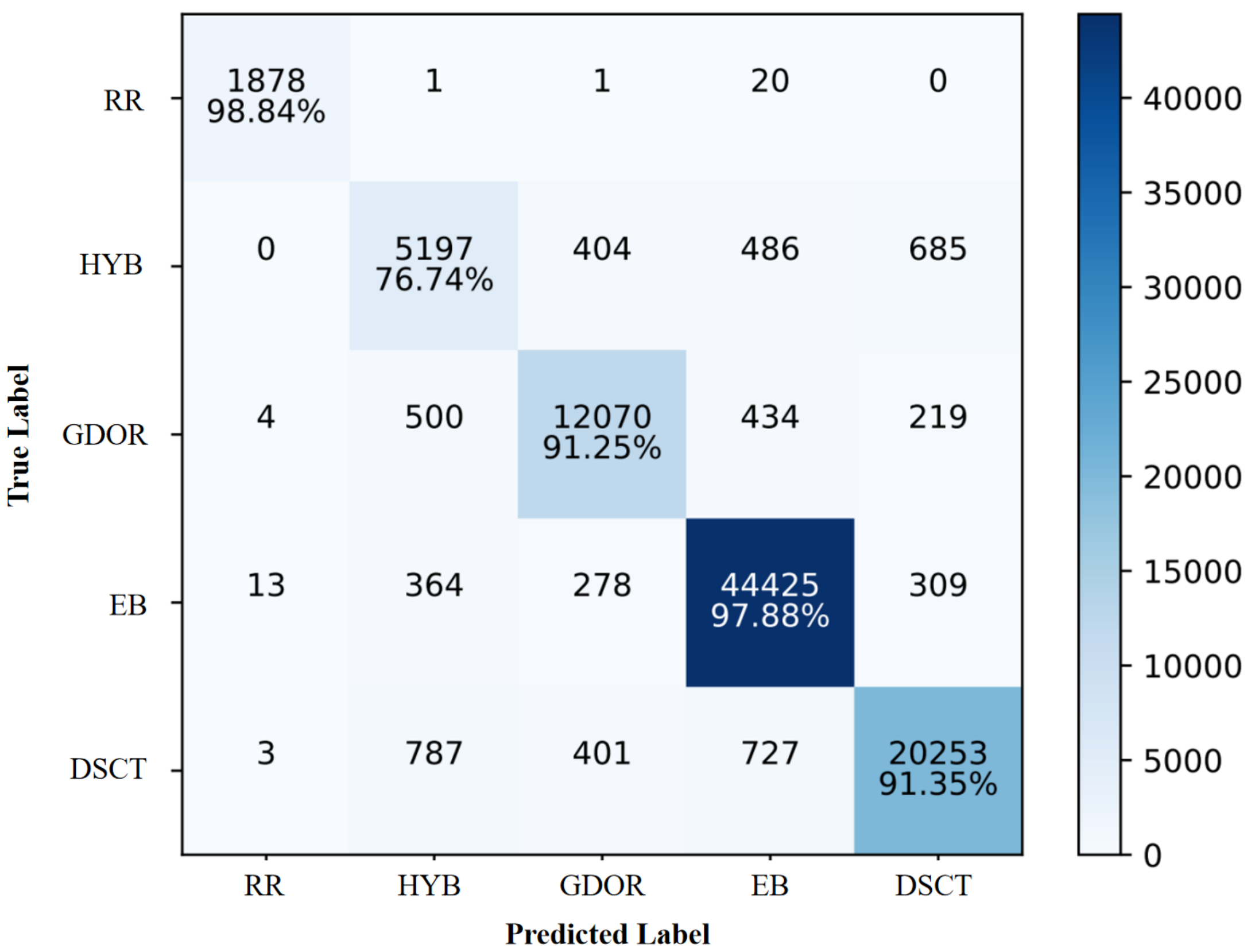}
\caption{The confusion matrix of the Conv1D + GRU model. The diagonal elements represent the correct predictions, along with the percentages of correctly classified observations for each true class.
\label{fig:gru}}
\end{figure*}

\begin{figure*}[ht!]
\centering
\includegraphics[width=0.9\textwidth]{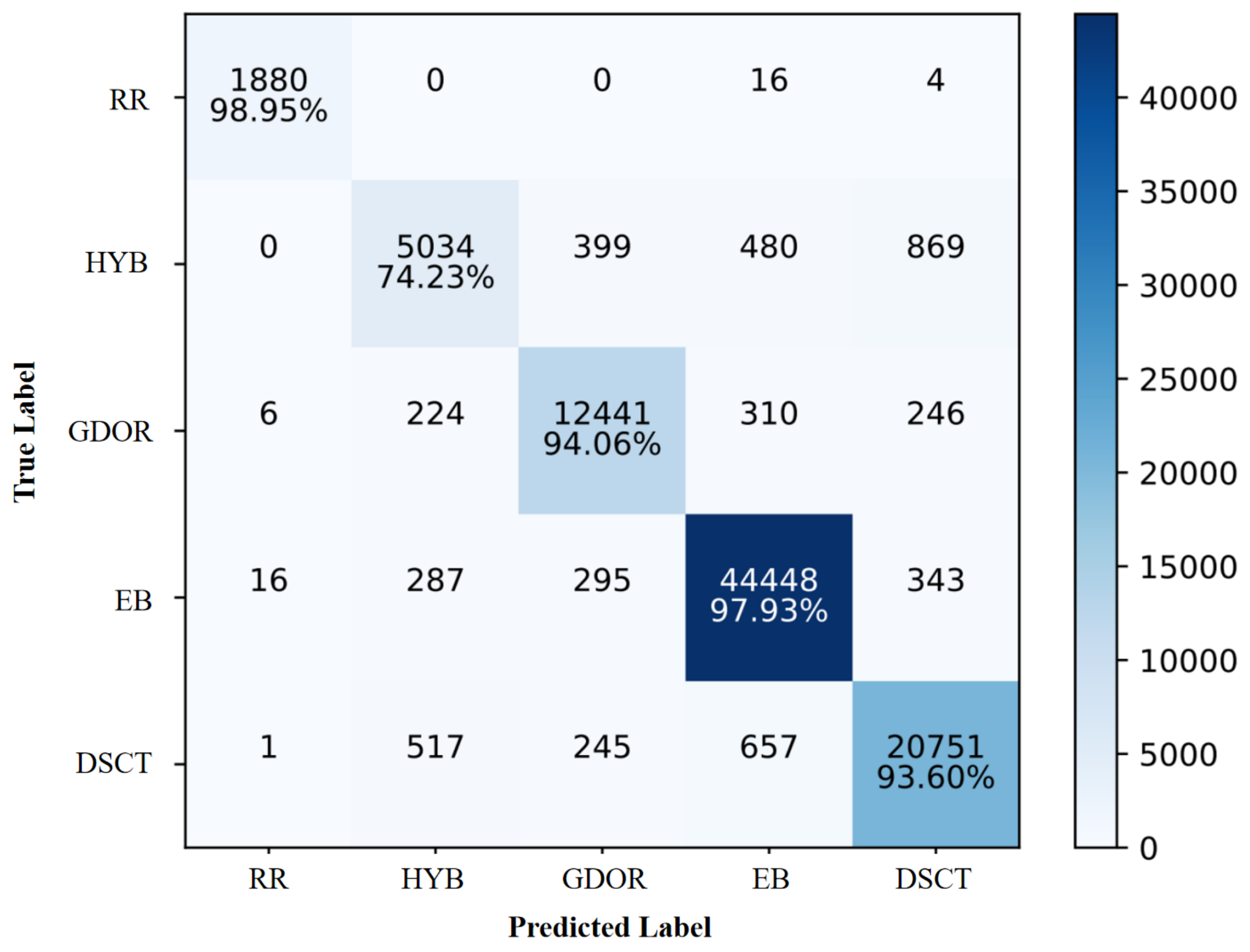}
\caption{The confusion matrix of the Conv1D + BiLSTM model. The diagonal elements represent the correct predictions, along with the percentages of correctly classified observations for each true class.
\label{fig:lstm}}
\end{figure*}

\begin{figure*}[ht!]
\centering
\includegraphics[width=0.9\textwidth]{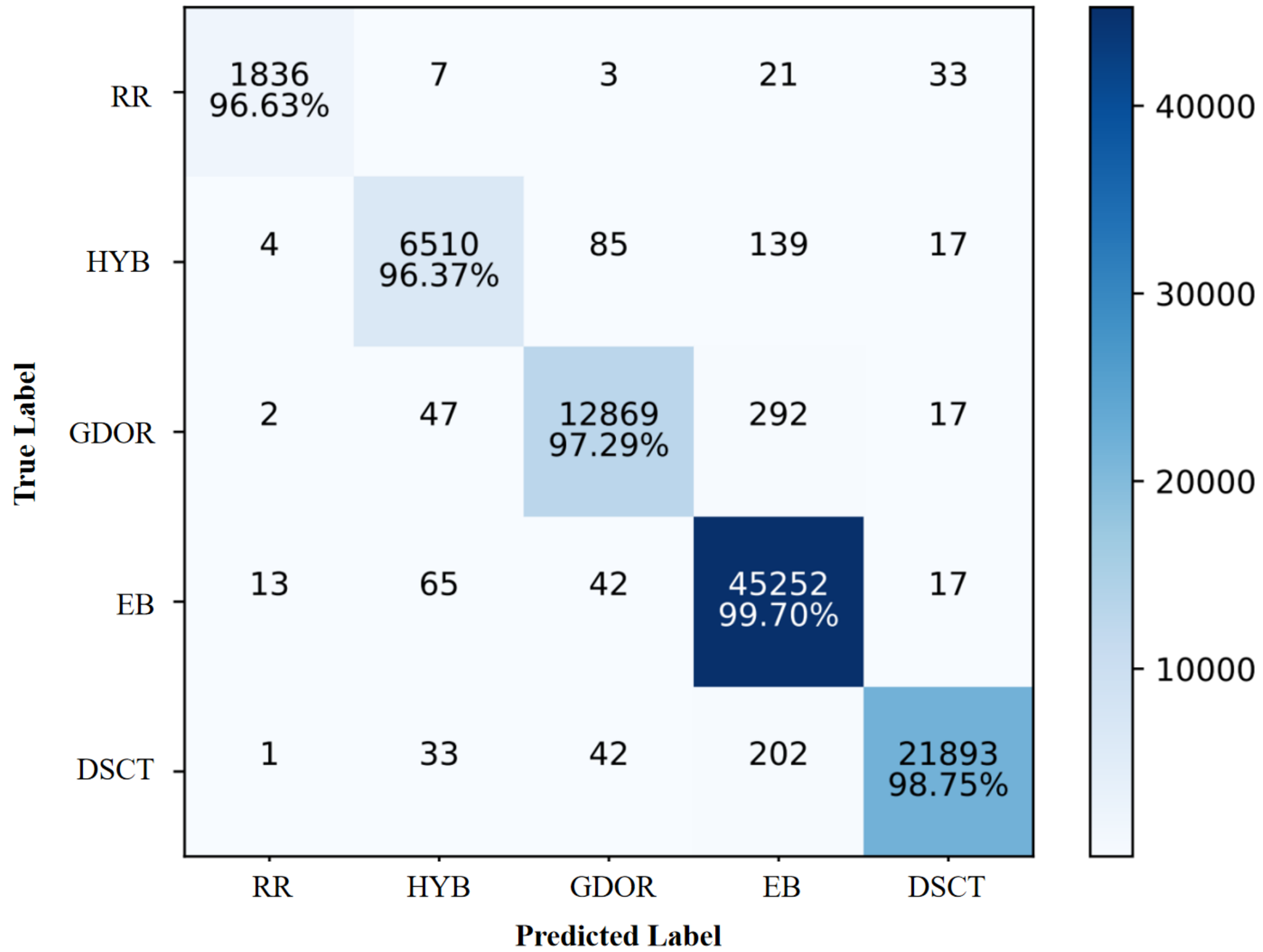}
\caption{The confusion matrix of the EfficientNet model. The diagonal elements represent the correct predictions, along with the percentages of correctly classified observations for each true class.
\label{fig:efficientnet}}
\end{figure*}

\begin{figure*}[ht!]
\centering
\includegraphics[width=0.9\textwidth]{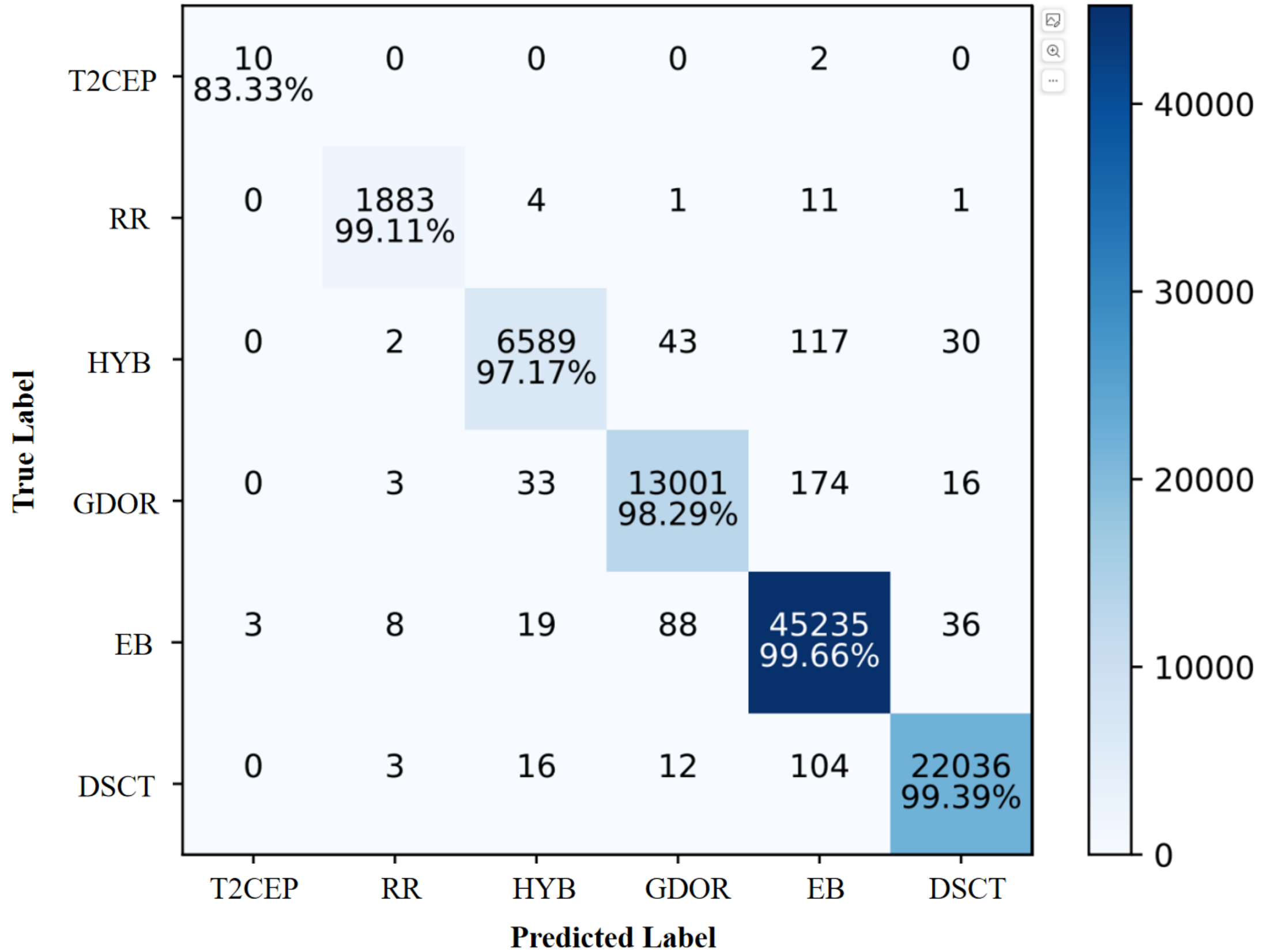}
\caption{The confusion matrix of the Swin Transformer model are presented, including the T2CEP class. The diagonal elements represent the correct predictions, along with the percentages of correctly classified observations for each true class.
\label{fig:swin}}
\end{figure*}

\begin{figure*}[ht!]
\centering
\includegraphics[width=\textwidth]{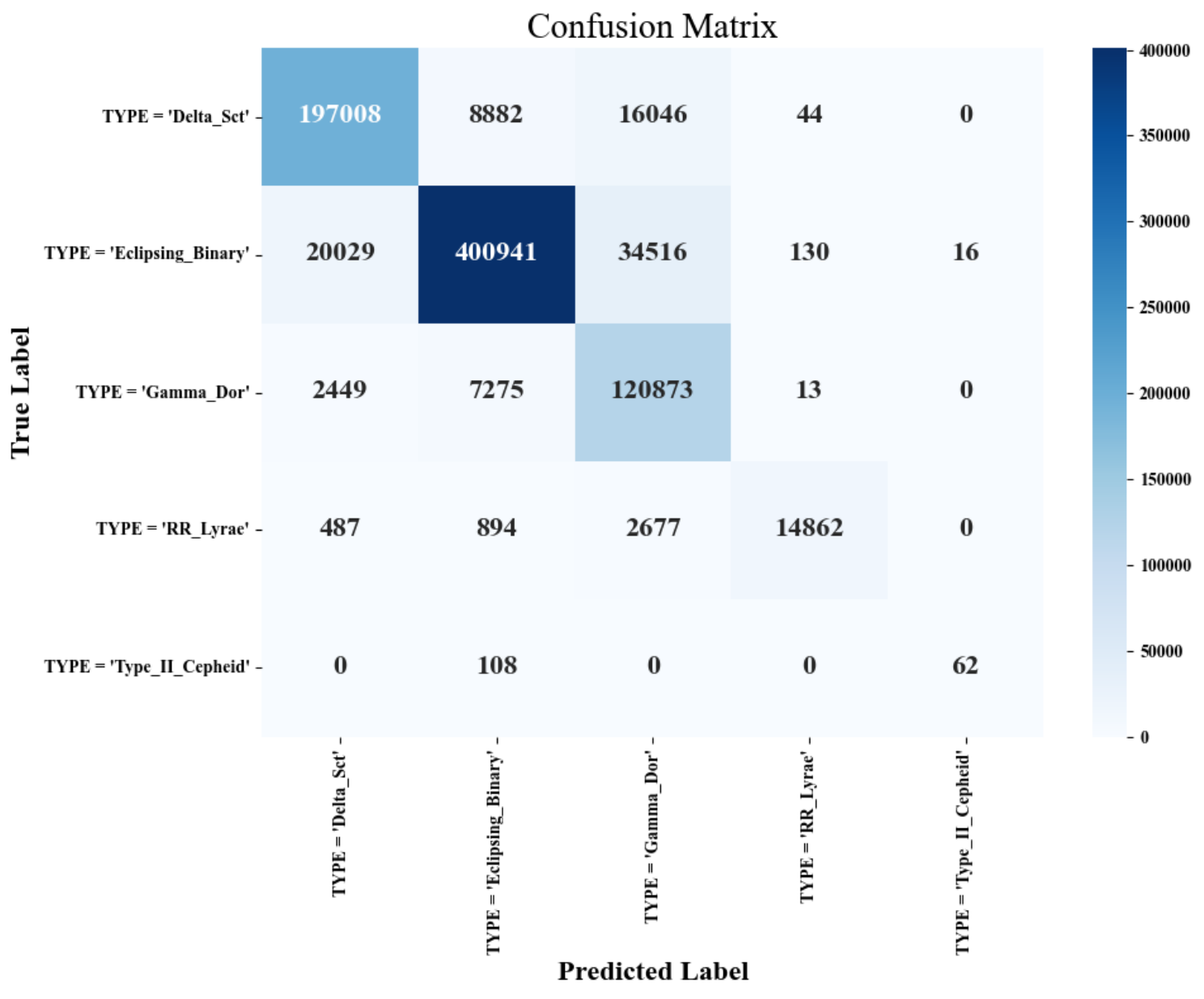}
\caption{The confusion matrix of the LLM method in StarWhisper LC.
\label{fig:LLMCM}}
\end{figure*}

\begin{figure*}[ht!]
\centering
\includegraphics[width=\textwidth]{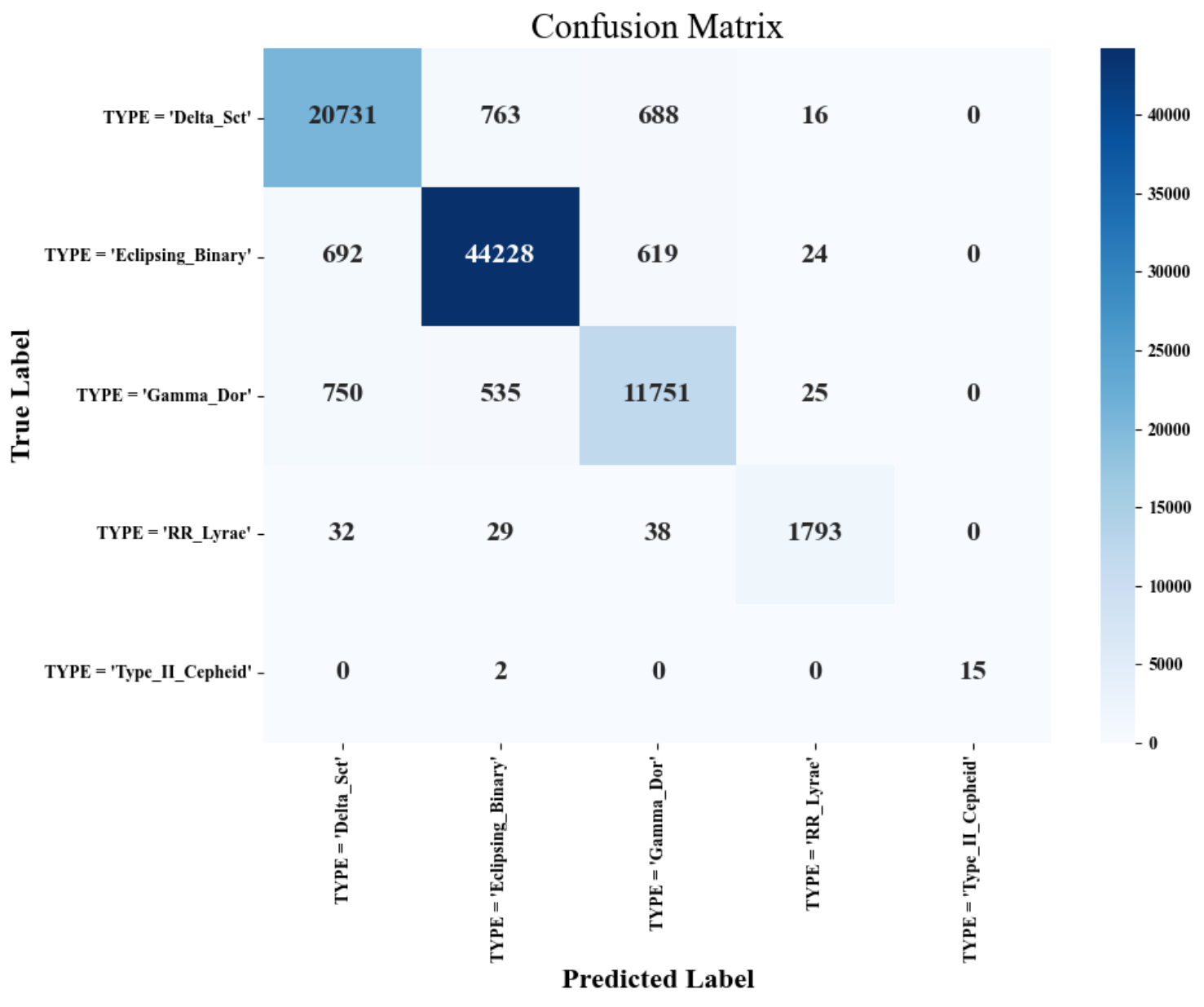}
\caption{The confusion matrix of the MLLM method in StarWhisper LC.
\label{fig:MLLMCM}}
\end{figure*}

\begin{figure*}[ht!]
\centering
\includegraphics[width=\textwidth]{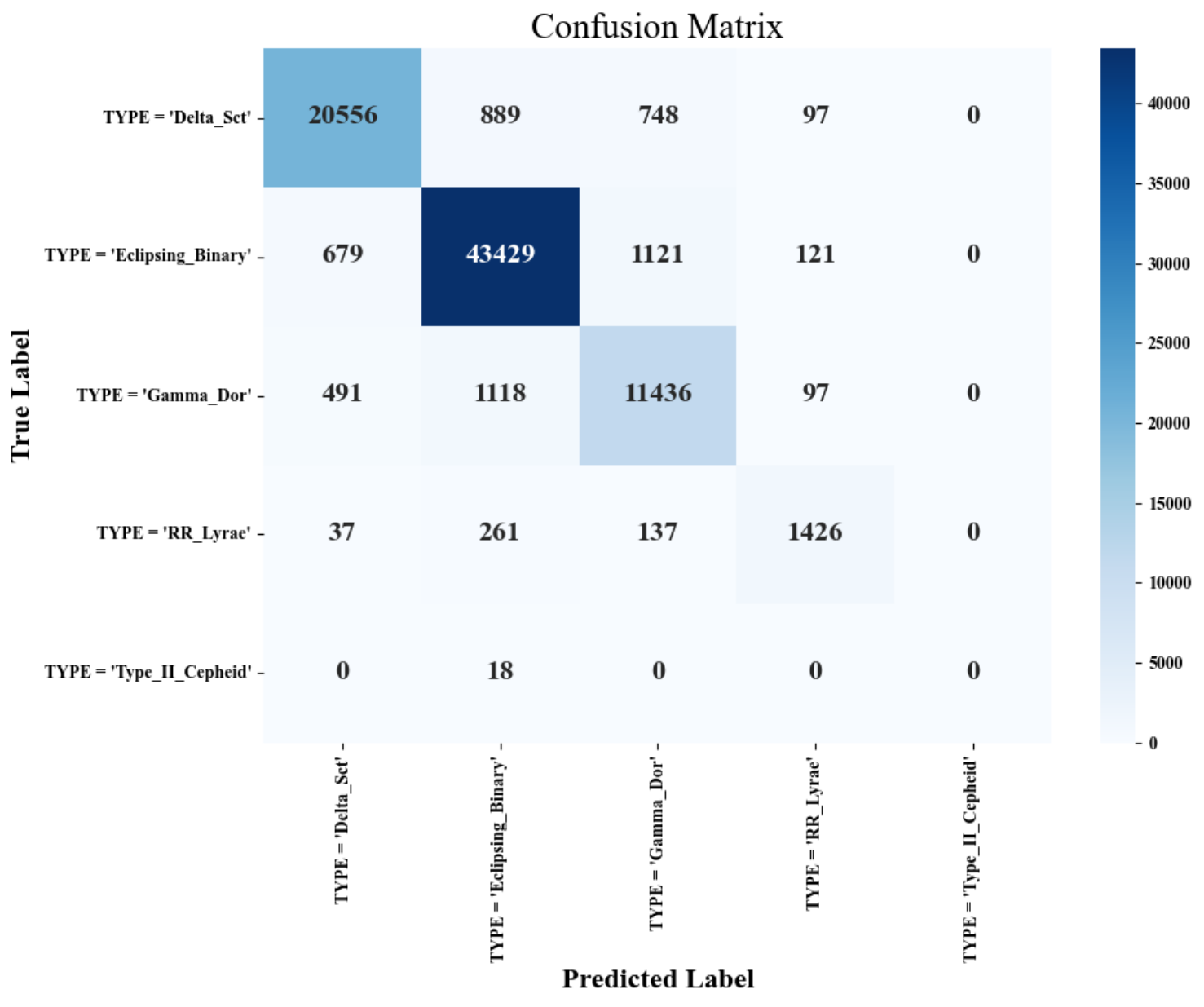}
\caption{The confusion matrix of the LALM mathod in StarWhisper LC.
\label{fig:LALM}}
\end{figure*}

In Figs. \ref{fig:transformer} to \ref{fig:LALM}, we present the confusion matrices of different models. The main body of each matrix, using varying shades of blue, illustrates the number of crossed objects in the classification model.

\end{document}